\documentclass{aa}
\usepackage{natbib}
\bibpunct{(}{)}{;}{a}{}{,}
\usepackage{units}
\usepackage{xstring}
\usepackage{multirow}
\usepackage{floatrow}
\usepackage{float}
\floatstyle{plaintop}
\restylefloat{table}
\usepackage{arydshln}

\usepackage{graphicx}
\usepackage[varg]{txfonts}
\usepackage[breaklinks]{hyperref}
\usepackage{microtype} 
\begin{document} 

   \title{
   Reliable detection and characterization of low-frequency polarized sources in the LOFAR M~51 field
   }


   \author{A. Neld\inst{1} 
   \and C. Horellou\inst{1} 
   \and D.D. Mulcahy\inst{2,}\inst{3}
   \and R. Beck\inst{3}
   \and S. Bourke\inst{1}
   \and T.D. Carozzi\inst{1}
   \and K.T. Chy\.zy\inst{4} 
   \and J.E. Conway\inst{1} 
   \and J.S. Farnes\inst{5} 
   \and A. Fletcher\inst{6}
   \and M. Haverkorn\inst{5}
   \and G. Heald\inst{7,}\inst{9}
   \and A. Horneffer\inst{3}
   \and B. Nikiel-Wroczy\'nski\inst{4}
   \and R. Paladino\inst{8}
   \and S.S. Sridhar\inst{9} 
   \and C.L. Van Eck\inst{5}
   }

    \institute{Department of Space, Earth and Environment, 
    Chalmers University of Technology,
    Onsala Space Observatory, 
    SE-439 92 Onsala, Sweden
    \email{cathy.horellou@chalmers.se}
     \and 
    Jodrell Bank Centre for Astrophysics, Alan Turing Building, School of Physics and Astronomy, The University of Manchester, Oxford Road, Manchester, M13 9PL, U.K.
     \and 
    Max-Planck-Institut für Radioastronomie, Auf dem Hügel 69, 53121, Bonn, Germany
     \and 
    Astronomical Observatory, Jagiellonian University, ul. Orla 171, 30-244 Krak\'ow, Poland
    \and 
    Department of Astrophysics/IMAPP, Radboud University, PO Box 9010, 6500 GL, Nijmegen, The Netherlands
    \and 
    School of Mathematics and Statistics, Newcastle University, Newcastle upon Tyne NE1 7RU, U.K.
    \and 
    CSIRO Astronomy and Space Science, 26 Dick Perry Ave, Kensington, WA 6151, Australia
    \and 
    INAF-Osservatorio di Radioastronomia, Via P. Gobetti, 101 I-40129 Bologna, Italy
    \and 
    Kapteyn Astronomical Institute, University of Groningen, Postbus 800, 9700 AV Groningen, The Netherlands
             }
   \date{Received 24 October 2017/ Accepted 12 June 2018}

 
  \abstract
   {The new generation of broad-band radio continuum surveys will provide large data sets with polarization information. New algorithms need to be developed to extract reliable catalogs of linearly polarized sources that can be used to characterize those sources and produce a dense rotation measure (RM) grid to probe magneto-ionized structures along the line of sight via Faraday rotation.}
   {The aim of the paper is to develop a computationally efficient and rigorously defined source-finding algorithm for linearly polarized sources.}
   {We used a calibrated data set from the LOw Frequency ARray (LOFAR) at 150~MHz centered on the nearby galaxy M51 to search for polarized background sources. 
   With a new imaging software, we re-imaged the field at a resolution of $18''\times15''$ and 
   cataloged a total of about 3000 continuum sources within 2.5$^\circ$ of the center of M~51. 
   We made small Stokes $Q$ and $U$ images centered on each source brighter than 100~mJy in total intensity (201 sources) and used RM synthesis to create corresponding Faraday cubes that were analyzed individually. For each source, the noise distribution function was determined from a subset of the measurements at high Faraday depths where no polarization is expected; the peaks in polarized intensity in the Faraday spectrum were identified and the $p$-value of each source was calculated. Finally, the false discovery rate method was applied to the list of $p$-values to produce a list of polarized sources and quantify the reliability of the detections.   
   We also analyzed sources fainter than 100~mJy but that were reported as polarized in the literature at at least another radio frequency. 
   }
   {Of the 201 sources that were searched for polarization, six polarized sources were detected confidently (with a false discovery rate of 5\%). 
    This corresponds to a number density of one polarized source per 3.3 square degrees, 
   or 0.3 source per square degree.
   Increasing the false discovery rate to 50\% yields 19 sources. A majority of the sources have a morphology that is indicative of them being double-lobed radio galaxies, and the ones with literature redshift measurements have $0.5<z<1.0$.
   }
   {We find that this method is effective in identifying polarized sources, and is well suited for LOFAR observations. 
   In the future, we intend to develop it further and apply it to larger data sets such as the LOFAR Two-meter Survey of the whole northern sky, LOTSS, and the ongoing deep LOFAR observations of the GOODS-North field.} 


   \maketitle

\section{Introduction}

One of the science drivers of the future Square Kilometre Array (SKA) is to produce a dense grid of polarized radio sources that could be used as background lights to probe magnetized media along their lines of sight in structures of various scales 
\citep[e.g.][]{2004NewAR..48.1289B, 
2015aska.confE.103G, 2015aska.confE..92J, 2016A&A...591A..13V}. 
The key effect is Faraday rotation, a birefrigence effect that causes the polarization angle of the linearly polarized wave emitted by a source to rotate as it propagates through a magneto-ionized medium: 
\begin{equation}
\chi = \chi_0 + {\rm RM} \lambda^2 \, , 
\end{equation}
where $\chi$ is the polarization angle measured at the wavelength of observation, $\lambda$, 
$\chi_0$ is the polarization angle of the emitted wave, 
and RM is the rotation measure. 
In the simple case when Faraday rotation occurs in a non-emitting foreground medium, the value of RM is equal to that of the Faraday depth of the source, 
$\phi(L)$, where 
$L$ is the entire pathlength to the source and $\phi(r)$
is 
a physical quantity which 
is proportional to the integral along the line of sight, $\ell$, 
 of the density of thermal electrons, $n_e$, times the magnetic field component parallel to the line of sight, $B_{\parallel}$:
\begin{equation}
\left( \frac{\phi(r)}{{\rm rad~m}^{-2}} \right) 
= 0.812 \int_{r}^{\rm observer} 
\left( \frac{n_e (\ell)}{{\rm cm}^{-3}} \right)
\left( \frac{B_{\parallel}(\ell)}{\mu{\rm G}} \right)
\left( \frac{d\ell}{\rm pc}  \right) 
\, . 
\label{eqphi}
\end{equation}

Rotation measures of polarized radio sources have been used to obtain information on magnetic fields in our own galaxy  
\citep[e.g.][]{2001ApJ...563L..31B, 
2007ApJ...663..258B}, 
in nearby galaxies
\citep[e.g.][]{1998A&A...335.1117H,  
2005Sci...307.1610G,  
2017MNRAS.467.1776K},  
in clusters of galaxies \citep[e.g.][]{2010A&A...513A..30B}, 
and to probe high-redshift absorbers \citep[e.g.][]{2013ApJ...772L..28B, 2014ApJ...795...63F}. 
A high number density of background polarized sources is an obvious requirement for such studies 
\citep[e.g.][]{2008A&A...480...45S}. 
Observations of a nine-square-degree field centered on the Andromeda galaxy led to the detection of 33 polarized sources at 350~MHz; fractional polarizations and RM's could be determined for those sources, but a larger catalog would be required to constrain the magnetic field pattern in M31 \citep{2013A&A...559A..27G}. 
The largest RM catalog available so far is based on the NVSS (NRAO VLA Sky Survey, \citealt{NVSS}) that covers the entire sky north of $-40^\circ$ declination at 1.4~GHz; it contains about 40\,000 sources, one polarized source per square degree \citep{taylor}. 
A significant unknown is the number density of polarized sources at low flux densities (sub-mJy; 
\citealt{2014ApJ...785...45R}) 
and at low frequencies, where depolarization effects are more significant 
\citep[e.g.][]{2011AJ....141..191F}. 
Much work is ongoing to produce larger catalogs of polarized sources and characterize their properties 
\citep[e.g.][]{2018A&A...613A..58V}. 

Following the formalism of \citet{burn},  
the \emph{observed} complex polarization $\mathcal{P}(\lambda^2) = Q(\lambda^2) + {\rm i} U(\lambda^2)$ can be expressed as
the integral over all Faraday depths of
the complex Faraday dispersion function\footnote{In this paper we call the {\it Faraday dispersion function} the complex-valued function denoted $\mathcal{F}$ and obtained from Eq.~(\ref{eqF}) where the integration is continuous and goes from $-\infty$ to $+\infty$; we denote $F$ the reconstructed $\mathcal{F}$ obtained from applying RM synthesis to a discrete set of measurements at defined frequencies and call it {\it a Faraday spectrum}.}
$\mathcal{F}(\phi)$, modulated by the Faraday rotation: 
\begin{equation}
\mathcal{P}(\lambda^2) 
= \int_{-\infty}^{+\infty} 
\mathcal{F}(\phi) e^{2{\rm i}\phi \lambda^2} 
d\phi \,. 
\label{eqpol}
\end{equation}
Equation~\ref{eqpol} is a Fourier-transform type relationship that can, in principle, be inverted to obtain $\mathcal{F}(\phi)$:
\begin{equation}  
\mathcal{F}(\phi) = 
\frac{1}{\pi}
\int_{-\infty}^{+\infty} 
\mathcal{P}(\lambda^2) 
e^{-2{\rm i}\phi \lambda^2} 
d\lambda^2  \, . \\ 
\label{eqF} 
\end{equation}

In practice, $\mathcal{F}(\phi)$ has to be reconstructed from a finite number of measurements at discrete frequencies. 
The RM synthesis method proposed by \citet{brentjens_bruyn} can be implemented efficiently and is commonly used to analyze polarization data, 
sometimes in combination with 
direct $q(\lambda^2)$ and $u(\lambda^2)$ fitting \citep[e.g.][]{mao}), 
where $q$ and $u$ are the $Q$ and $U$ Stokes parameters normalised to the total intensity $I$ 
at wavelength $\lambda$. 
While RM synthesis is well suited for single (and strong) Faraday depth components, it has difficulty recovering multiple and complex components \citep[e.g.][]{2012MNRAS.421.3300O, 
2015ApJ...815...49A, 
2018MNRAS.474..300S}
and it has been shown that the uncertainties on the derived RM are not accurate for sources with non-zero spectral indices (\citealt{2017MNRAS.466..378S}, 
\citeyear{2018MNRAS.473.3732S}). 
Efficient and reliable source-finding algorithms need to be developed in order to analyze the large amount of data that upcoming radio surveys will deliver. 
Several packages are available to identify radio continuum sources in total intensity (see \citet{2012MNRAS.422.1812H} for a review). 
For several reasons, those algorithms may not provide correct results when applied to polarization data. 
One of these reasons is the non-Gaussian nature of the noise in polarized intensity, $P$: 
the noise in $P$ follows a Rician distribution
in the case of Gaussian noise in Stokes $Q$ and $U$ (with zero mean and same variance).
Methods have been developed to correct for the bias introduced by the non-Gaussianity in $P$ \citep[e.g.][]{1974ApJ...194..249W,2017A&A...600A..63M}. 
However, the noise in $Q$ and $U$ may be non-Gaussian, which causes a significant increase in the false detection rates when detection thresholds based on predefined signal-to-noise ratios are used \citep{non_rician}. 

Another difficulty is the instrumental polarization that manifests itself as a leakage from Stokes $I$ into Stokes $Q$ and $U$ and contaminates the measurements in the entire frequency band, 
and in both on-source and off-source regions of the $Q$ and $U$ images. This means that $Q$- and $U$-based detection methods (such as the analytic method by \citet{2012MNRAS.424.2160H}) 
are not directly applicable to LOFAR data and the analysis must be done in Faraday space where the instrumental polarization effects are concentrated to a region near Faraday depth $\phi = 0$.

Recently, \citet{2018MNRAS.474.3280F} 
proposed a computationally efficient source-finding algorithm that makes use of so-called Faraday moments (moments of the $Q$, $U$, and $P$ distributions: mean, standard deviation, skewness and excess kurtosis). The approach is easy to understand intuitively as a high polarization would produce a high mean in $Q$ and/or $U$, and a high RM a high standard deviation in $Q$ and $U$. 
However, the method provides a source list that, although complete, contains a large number of false detections due to instrumental polarization and needs to be followed up with RM synthesis to eliminate the unreliable sources. 

Since the amount of Faraday rotation is proportional to $\lambda^2$, it is of particular interest to observe at long wavelengths (low frequencies) and over a very broad frequency range to obtain more precise  rotation measures. The LOw Frequency ARray (LOFAR, \citealt{lofar}) operates in two frequency ranges: 30 -- 80 MHz with the Low-Band Antennas (LBA) and 120 -- 240 MHz with the HBA. For this work, low-frequency HBA data (up to about 180~MHz) were used as they offer greater and more uniform sensitivity as a function of frequency. Additionally, the process of data calibration is facilitated due to both the higher sensitivity and the fact that ionospheric effects are less severe in the higher band. 
LOFAR is equipped with receivers and correlators that allow observations across a large instantaneous bandwidth with a great number of frequency channels. This new instrumentation results in a significant boost in sensitivity. In addition, the large field of view of LOFAR makes it an efficient survey instrument
\citep[e.g.][]{2015A&A...582A.123H, 
2017A&A...598A.104S}. 

Polarization work with LOFAR has been very challenging so far because of ionospheric Faraday rotation 
\citep{2013A&A...552A..58S}, 
instrumental polarization, uncertainty in the primary beam model, and the generally strong Faraday depolarization at low frequencies 
\citep[e.g.][]{1998MNRAS.299..189S,1999MNRAS.303..207S}. 
Calibration and imaging at high resolution ($\leq 1'$) is hard at low frequencies, and so beam depolarization can often be a limitation.
Despite these difficulties, polarization studies 
are now becoming possible as the nature of the data and the characteristics of the instrument become better understood. This is also important for investigations of the epoch of reionization (EoR), since polarization leakage may mimic an EoR signal \citep{2016MNRAS.462.4482A}. 

Diffuse Galactic foreground polarization has been detected by LOFAR 
in deep fields (the ELAIS N1 field, \citealt{2014A&A...568A.101J};  
the  3C~196 field, \citealt{2015A&A...583A.137J}), 
and in the Galactic foreground of the nearby galaxy IC~342 
\citep{2017A&A...597A..98V}. 
The Murchison Widefield Array (MWA) also detects diffuse Galactic polarization with better sensitivity to the largest scale emission \citep{2016ApJ...830...38L}, 
but relatively few extragalactic sources so far in polarization \citep{2013ApJ...771..105B, 2017PASA...34...40L}. 
LOFAR provides higher angular resolution and sensitivity and thereby the potential to probe the fainter  source population.
Polarization was detected in the lobe of a radio galaxy \citep{2015A&A...584A.112O}. 
No diffuse polarization was found toward the nearby spiral galaxy M51, but six background polarized sources were detected in the M51 field \citep{mulcahy}.
\citet{2018MNRAS.474.3280F} 
applied their Faraday moments method to the LOFAR data of the M51 field. 
Recently, \citet{2018A&A...613A..58V} 
developed a pipeline to search for polarization in regions of the sky covered by the LOFAR Two-Meter Sky Survey (LOTSS, \citealt{2017A&A...598A.104S}). 
This work resulted in a catalog of 92 
polarized sources at 150~MHz in an area of 570 square degrees, corresponding to a density of one source per 6.2 square degrees. 
The data were imaged at low angular resolution (4$'$) and were strongly affected by polarized foregrounds, so it is likely that the detection rate of polarized sources would increase at higher angular resolution. 

\begin{figure*} 
\centering
\includegraphics[width=17.7cm]{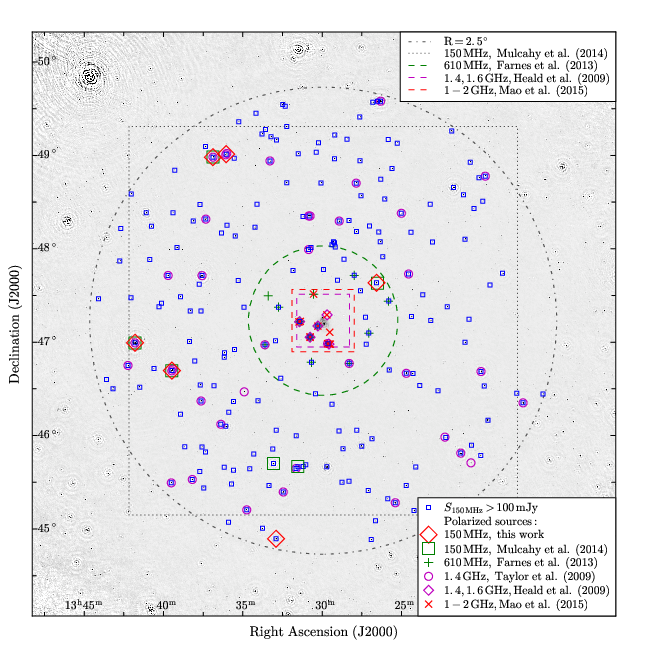}
\caption{LOFAR 150~MHz image of the field centered on nearby galaxy M51. The size of the synthesized beam is $\unit[18]{''}\times\unit[15]{''}$. Roughly 3\,000 radio continuum sources were detected in this image within $2.5^\circ$ of the center of M51 (dotted-dashed black circle). 
We searched for polarization in all sources brighter than 100~mJy (201 sources; small blue squares). 
The locations of the sources that were found to be polarized in this work and/or in other studies are also indicated. 
The entire field is included in the 1.4~GHz polarization catalog of \citet{taylor}.
The regions examined in other studies are shown as the 
dotted black square (LOFAR 150~MHz, \citealt{mulcahy}), 
dashed green circle (GMRT 610~MHz, \citealt[][approximate]{farnes}), 
dashed magenta square (WSRT 1.4 and 1.6~GHz, \citealt{heald}), 
dashed red square (VLA 1 -- 2~GHz, \citealt{mao}).
The FITS file of the LOFAR total intensity image of the field will be made available electronically.
}
\label{fig1} 
\end{figure*}

In this study, we used the calibrated LOFAR measurement sets of the M51 field published by \citet{mulcahy} to carry out a systematic search for polarized sources in the field. 
We re-imaged the field with an up-to-date LOFAR imager and developed a new algorithm to extract a catalog of polarized sources and quantify the rate of false detections. The method is entirely empirical and no assumption is made on the nature of the noise. 

The paper is organized as follows. 
The main characteristics of the data set are given in Sect.~\ref{sectObs}. 
The analysis of the continuum data is presented in Sect.~\ref{sectConti} 
and that of the polarization in Sect.~\ref{sectPolar}. 
The results are presented in Sect.~\ref{sectResults}. 
In Sect.~\ref{sectDiscussion}, the sources that are most confidently detected in polarization are discussed individually and the LOFAR measurements are compared to other available radio polarization measurements. 
The method used to identify the polarized sources is compared to the more standard methods based on a fixed signal-to-noise ratio.  Depolarization effects and the insensitivity of the observations to Faraday-thick structures are discussed. 
Finally, we conclude in Sect.~\ref{sectConclusion}.

\section{The LOFAR data}
\label{sectObs}

The M51 field was observed in 2013 for eight hours using the LOFAR HBA\footnote{Proposal LCO\_043, PI R. Beck}. During the observation, the field was never below $40^\circ$ elevation. This is important as simulations have shown that LOFAR's sensitivity to polarization is significantly reduced at low elevations (T. Carozzi, private communication).
There were eight frequency blocks, each approximately $\unit[6]{MHz}$ wide, spread evenly between $\unit[115]{MHz}$ and $\unit[175]{MHz}$. 
In total there were 1952 frequency channels with a channel width of $\unit[24.4]{kHz}$.
3C\,295 was used for flux and initial phase calibration. We estimate a 10\% calibration error in the total intensity flux. Due to the difficulty in calibrating polarization with LOFAR, we cannot estimate the calibration error in polarized intensity confidently. More details about the observation and calibration are available in \citet{mulcahy}. 

\section{Analysis of the continuum data}
\label{sectConti}

\subsection{Imaging}
\label{imaging_sect}

The field was imaged in total intensity using \textsc{wsclean 2.2}\footnote{{\tt https://sourceforge.net/projects/wsclean}} \citep{wsclean}. 
We imaged a square of $6.25^\circ \times 6.25^\circ$ centered on M\,51, with a $18''\times15''$ elliptical beam 
and a pixel size of 5$''$. 
We used Briggs weighting \citep{1995AAS...18711202B} with a robustness parameter of 0. 
The image was cleaned down to $3\sigma$, after which a mask was applied with \textsc{wsclean}'s auto-masking option and the image was cleaned to the $0.3 \sigma$-level, as recommended in the {\sc wsclean} documentation. 
All frequency channels were imaged individually and those strongly affected by radio frequency interference were discarded, including the whole last block. This left 1694 channels with a maximum frequency of $\unit[168]{MHz}$.

Figure~\ref{fig1} is an image of the field obtained after differential beam correction. 
The data that we used had already been corrected for the response of the LOFAR primary beam, calculated at the phase center \citep{mulcahy}; 
we applied the differential beam correction in \textsc{wsclean}\footnote{The differential beam was applied using the {\tt wsclean} flags {\tt -apply-primary-beam} and {\tt -use-differential-lofar-beam}} 
based on the so-called Hamaker model\footnote{
Hamaker J.~P., 2011, Tech. Rep., Mathematical-Physical Analysis of the
Generic Dual-dipole Antenna. ASTRON, Dwingeloo (H11)
}
(for more information, see e.g. Sect.~2.2.2 of \citet{2015MNRAS.451.3709A} and references therein). 

The primary beam correction and phase errors cause the noise to vary across the image. Across the inner region of $2.5^\circ$ radius the RMS noise in the full-bandwidth Stokes $I$ image varies from $\unit[200]{\mu Jy\, beam^{-1}}$ to $\unit[800]{\mu Jy\, beam^{-1}}$, depending on distance from the phase center and proximity to bright sources.

We also produced 
full-bandwidth $Q$ and $U$ images using the same parameters as for the $I$ image (but without cleaning, due to the low signal-to-noise ratio). 
The noise in the $Q$ and $U$ images is not as affected by nearby sources; it varies mostly with distance from the phase center within the primary beam. 
It varied from $\unit[100]{\mu Jy\, beam^{-1}}$ at the center to $\unit[200]{\mu Jy\, beam^{-1}}$ at a distance of $2.5^\circ$ from the center of M\,51.

The characteristics of the full-bandwidth $I$ image are given in Table~\ref{tab1}. The table also lists the number of sources detected in the field, as discussed in the following Section. 

\subsection{Source identification}

   \begin{table}
      \caption[]{\label{tab1} Characteristics of the imaging and field.}
 \begin{tabular}{l l}
            \hline
            \hline
            \noalign{\smallskip}
            Synthesized beam                & $18''\times15''$\\
            Beam position angle           & $104^\circ$\\
            $\sigma_I$                      &200 -- 800\,$\mu$Jy\,beam$^{-1}$\\
            $N_{\rm LOFAR}(R < 2.5^\circ)$  &\textasciitilde{}3,000\tablefootmark{a}\\
            $N_{\rm TGSS} (R < 2.5^\circ)$     &324\tablefootmark{b}\\
            $N_{\rm LOFAR}(R < 2.5^\circ, S_{\rm 150\,MHz} > 100$\,mJy)    &201\tablefootmark{a}\\
            $N_{\rm Taylor}(R < 2.5^\circ)$  &38\tablefootmark{c}\\
            \noalign{\smallskip}
            \hline
   \end{tabular}
   \tablefoot{
    \tablefoottext{a}{This work.}
    \tablefoottext{b}{The first alternative data release TGSS ADR1 of \citet{TGSS_ADR1} at 150~MHz, 25$''$ resolution and noise level of about 5~mJy~beam$^{-1}$.}
    \tablefoottext{c}{Polarized sources in the \citet{taylor} catalog at 1.4~GHz and 45$''$ resolution.}
   }
   \end{table}

To identify the continuum sources in the field we used the Python Blob Detector and Source Finder, {\tt pyBDSF}\footnote{Formerly {\tt pyBDSM} \citep{pybdsm}. {\tt http://www.astron.nl/citt/pybdsf}}. 
We used a $250''$ box to calculate the RMS map\footnote{{\tt RMS\_box}=(50, 15)}, while the other parameters were kept at the default values.
This resulted in the detection of 3\,032 sources within $2.5^\circ$ of the center of M51, though a number of them (\textasciitilde{}10-20) were visually determined to be false detections from phase errors around strong sources. 

\begin{figure} 
\centering
\includegraphics[width=\hsize]{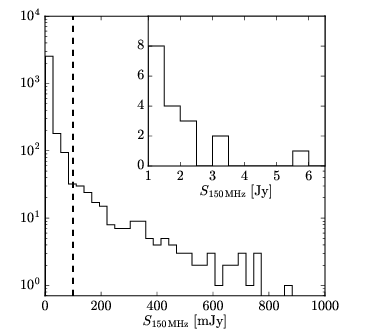}
\caption{Number counts of continuum sources detected within $2.5^\circ$ of the center of M~51. 
The vertical line shows the 100~mJy flux density threshold used in the polarization search. 
The inset shows the distribution for sources brighter than 1~Jy. 
}
\label{figNbercounts}
\end{figure}

Figure~\ref{figNbercounts} shows the corresponding number counts. The vertical dashed line indicates the 100~mJy flux density threshold used in the polarization search. 
The choice of this threshold is justified in Sect.~\ref{sectTheSample}.

This LOFAR catalog was cross-matched with the first alternative data release of the Tata Institute of Fundamental Research (TIFR) Giant Metrewave Radio Telescope (GMRT) Sky Survey (TGSS, hereafter TGSS ADR1, \citealt{TGSS_ADR1}) in the same region. All but three of the 324 TGSS ADR1 sources were found in our LOFAR catalog. 
All three undetected TGSS ADR1 sources were located near sidelobes of bright sources in the LOFAR image.  

\begin{figure} 
\includegraphics[width=8.8cm]{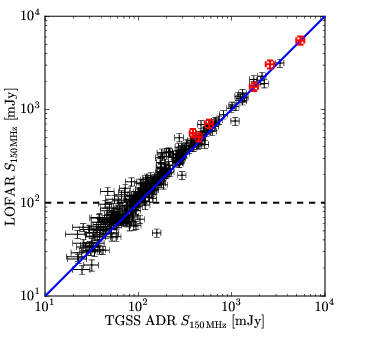}
\caption{Comparison of the flux densities measured in the LOFAR image and in the first alternative data release of the TGSS survey \citep{TGSS_ADR1}. 
The diagonal (solid line) is the 1:1 line. The red markers represent the six continuum sources in which polarization was most securely detected. 
The dashed line at $10^2$\,mJy indicates the flux density threshold used in the polarization search. 
}
\label{tgss_fig}
\end{figure}

Figure~\ref{tgss_fig} shows a comparison between the LOFAR flux density measurements and those in TGSS ADR1. The LOFAR flux densities are higher by 20\% on average, and this effect decreases with increasing flux density. This difference is too large to be only due to calibration error. An explanation might be that the higher sensitivity of LOFAR allows observation of diffuse emission that is not detected in the TGSS ADR1. 
Another explanation is that there is an increasing degree of incompleteness at low flux densities (because intrinsically faint sources are only seen at the center of the LOFAR image, whereas the bright sources are recovered at all radii). 
The faint end of the scatter in low-flux-density bins is truncated and we are left with a positive bias relative to TGSS which is mosaiced and has roughly uniform sensitivity across the survey area.

\section{Analysis of the polarization data}
\label{sectPolar}

\begin{figure} 
\includegraphics[width=8.45cm]{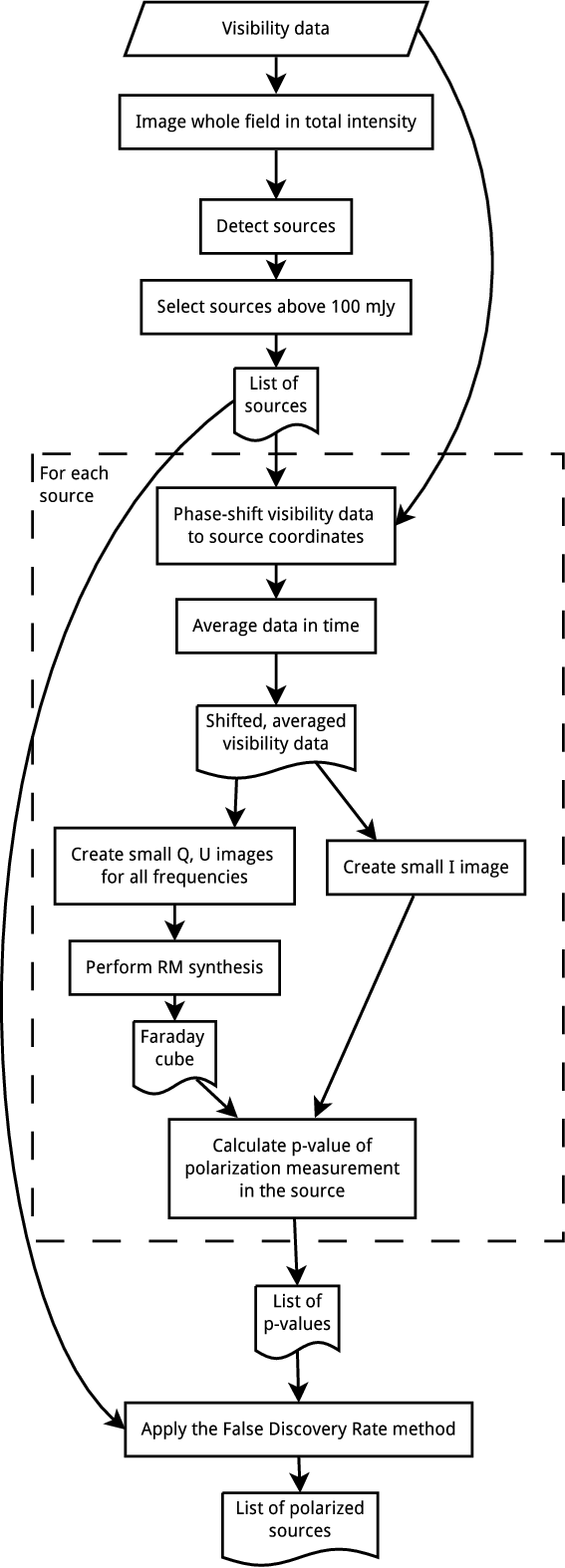}
\caption{Process used in this paper to obtain a list of polarized sources from visibility data 
(see Sect.~\ref{sectConti} and Sect.~\ref{sectPolar}).} 
\label{flowchart}
\end{figure}

A flowchart outlining the method is shown in Fig.~\ref{flowchart}.

\subsection{The sample} \label{sectTheSample}

We searched for polarization in all sources of the LOFAR 150~MHz image with a flux density greater than $\unit[100]{mJy}$ and located within $2.5^\circ$ of the center of M51 (201 sources, eight of which are not in the TGSS ADR1 catalog). 
The catalog is published electronically.

The 100~mJy flux density threshold was set on the basis of the noise level in the images, 
$\sigma_I < 0.8$~mJy~beam$^{-1}$ across the entire field of view (FOV), so that all sources brighter than 100~mJy would be detected at a signal-to-noise ratio (S/N) greater than 100. 
A polarized source with a fractional polarization of 1\% would be detectable at a S/N$ > 5$,  
since $\sigma_{Q,U} < 0.2$~mJy~beam$^{-1}$ over the FOV. Lower fractional polarizations would be detectable in brigher sources. 

We also examined six sources below this threshold that had been detected in polarization at other radio frequencies. 
These sources were imaged in Faraday space and were analyzed, but not included in the false discovery rate (FDR) analysis  described in Sect.~\ref{fdr}. 

The locations of all examined sources are indicated in Fig.~\ref{fig1}. 
The sources brighter than 100~mJy are distributed rather uniformly across the field. 
The six weaker sources are located in the central region, in the areas that were mapped in deep observations at higher frequencies by \citet{farnes}, \citet{heald}, or \citet{mao}.

\subsection{Creating Faraday cubes}
\label{making_cubes}

We imaged each source using the procedure described below.

First, we phase-shifted the $(u,v)$ data to the source location and averaged them in time to $\unit[140]{s}$, using {\tt DPPP}\footnote{Formerly {\tt NDPPP}, part of the standard LOFAR imaging pipeline \citep[e.g.][]{lofar_pipeline}.}.
    
Then we used {\sc wsclean} to create small images  ($\unit[4.3]{'}\times\unit[4.3]{'}$) of all four Stokes parameters centered on the source for all frequencies. The small image size made the high time averaging possible; the smearing that occurs when averaging in time is smaller near the phase center.
Only baselines shorter than $\unit[18\,000]{\lambda}$ were included to give all channel maps the same angular resolution. 
The potential intensity loss due to time smearing given the parameters here ($4.3'$ image, $15''$ beam, 140~s) is $< 1$\% \citep{1999ASPC..180..371B}. 
The channel maps were not cleaned because of the low signal-to-noise ratio in the individual $Q$, $U$ images. Briggs' weighting \citep{1995AAS...18711202B} was used, with a robustness parameter of 0.
The {\sc wsclean} differential primary beam correction\footnote{In {\sc wsclean} versions prior to $2.1$ the sign of Stokes $Q$ was wrong. As we used version $2.2$, this is not an issue.} was applied.
We also imaged the source in total intensity (combining all frequency channels). For this image, cleaning was performed in the same way as for the image of the whole field, as described in Sect.~\ref{imaging_sect}.

\begin{figure}
\includegraphics[width=8.8cm]{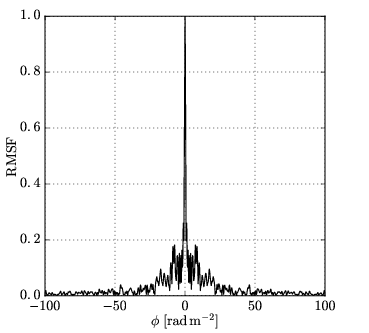}
\caption{Absolute value of the rotation measure spread function (RMSF) corresponding to the frequency coverage of the LOFAR data used in this work. The full-width half maximum of the RMSF is $\delta\phi \approx 0.96$~rad~m$^{-2}$. 
}
\label{rmsf} 
\end{figure}

Finally, we performed RM synthesis on the $Q$ and $U$ images using {\tt pyrmsynth}\footnote{{\tt https://github.com/mrbell/pyrmsynth}}. 
Faraday cubes were created between $\pm 500$~rad~m$^{-2}$  
and cleaned to reduce the sidelobes in Faraday space down to 3~$\sigma_{\rm F}$ 
(where $\sigma_{\rm F}$ is the standard deviation of the Faraday spectrum $|F(\phi)|$ at a given pixel in RA, Dec) 
with the {\tt RM-CLEAN} algorithm \citep{heald}. 
The rotation measure spread function (RMSF) is shown in Fig.~\ref{rmsf}. 
Since the noise was higher near the edges of the images and at high values of $|\phi|$, we used slightly smaller Faraday cubes for the analysis: 
\begin{equation}
3'\times3', |\phi| < 450\,{\rm rad~m}^{-2}\, . 
\end{equation}

The limits of RM synthesis given the frequency coverage of the data set can be calculated from equations 61-63 in \citet{brentjens_bruyn}:
\begin{gather}
    \delta\phi           \approx \unit[0.96]{rad\,m^{-2}} \\
    \Delta\phi_{\rm max} \approx \unit[0.99]{rad\,m^{-2}} \\
    |\phi_{\rm max}|     \approx \unit[1350]{rad\,m^{-2}}\, , 
\end{gather}

\noindent where $\delta\phi$ is the resolution in Faraday depth (strictly speaking the full-width half maximum of the RMSF), 
$\Delta\phi_{\rm max}$ is the largest scale in Faraday depth to which the data are sensitive, 
and $|\phi_{\rm max}|$ is the largest Faraday depth in absolute value that can be detected. 
Since $\Delta\phi_{\rm max}$ is barely larger than the resolution in Faraday depth, polarization will only appear as unresolved peaks in Faraday space.

\subsection{Faraday voxels, Faraday cells, and local maxima}

A Faraday voxel is a 3D pixel in the Faraday volume. Each voxel has a size of 
$2\arcsec\times2\arcsec\times\unit[0.2]{rad\,m^{-2}}$. 

A Faraday cube can be regarded as a number of independent resolution elements, which we will call {\emph {Faraday cells}}. 
The spatial component of each Faraday cell has the size of the synthesized beam and the 
third dimension is the resolution element in Faraday space: 
$18\arcsec\times15\arcsec\times\unit[0.96]{rad\,m^{-2}}$. 

Note that the Faraday cells are not rectangular parallelepipeds, but 3D Gaussians. 
Each imaged Faraday cube contained roughly 170\,000 such independent cells. 
In the analysis, individual cells are not used; the relevant quantity is the number of cells in a Faraday volume, as it is the number of independent measurements. 

Because the voxels in a Faraday volume are correlated due to oversampling, the analysis was performed on local maxima that were identified by examining the values of $|F|$ in adjacent voxels\footnote{The {\tt SciPy} routine {\tt ndimage.filters.maximum\_filter} was used for this.}. 
We assume that each local maximum corresponds to one cell.
The density of peaks at or above a given $F$ was obtained by dividing the number of identified local maxima by the number of cells in the considered Faraday volume. 

\subsection{Regions in the Faraday cubes}
\label{subsectRegions}

\begin{figure*} 
\centering
\floatbox[{\capbeside\thisfloatsetup{capbesideposition={right,center},capbesidewidth=3cm}}]{figure}[\FBwidth]
{\caption{Illustration of the different regions used in the analysis of a Faraday cube. 
The green cylinders represent the region searched for polarization (at Faraday depth $|\phi| < 100$~rad~m$^{-2}$ and outside the central region that is contaminated by instrumental effects). 
The outer regions (100 rad~m$^{-2} < |\phi| < 450$ rad~m$^{-2}$) 
were used to characterize the noise. 
The images have a size of $3'\times3'$.} \label{region_figure}}
{\includegraphics[width=14.5cm]{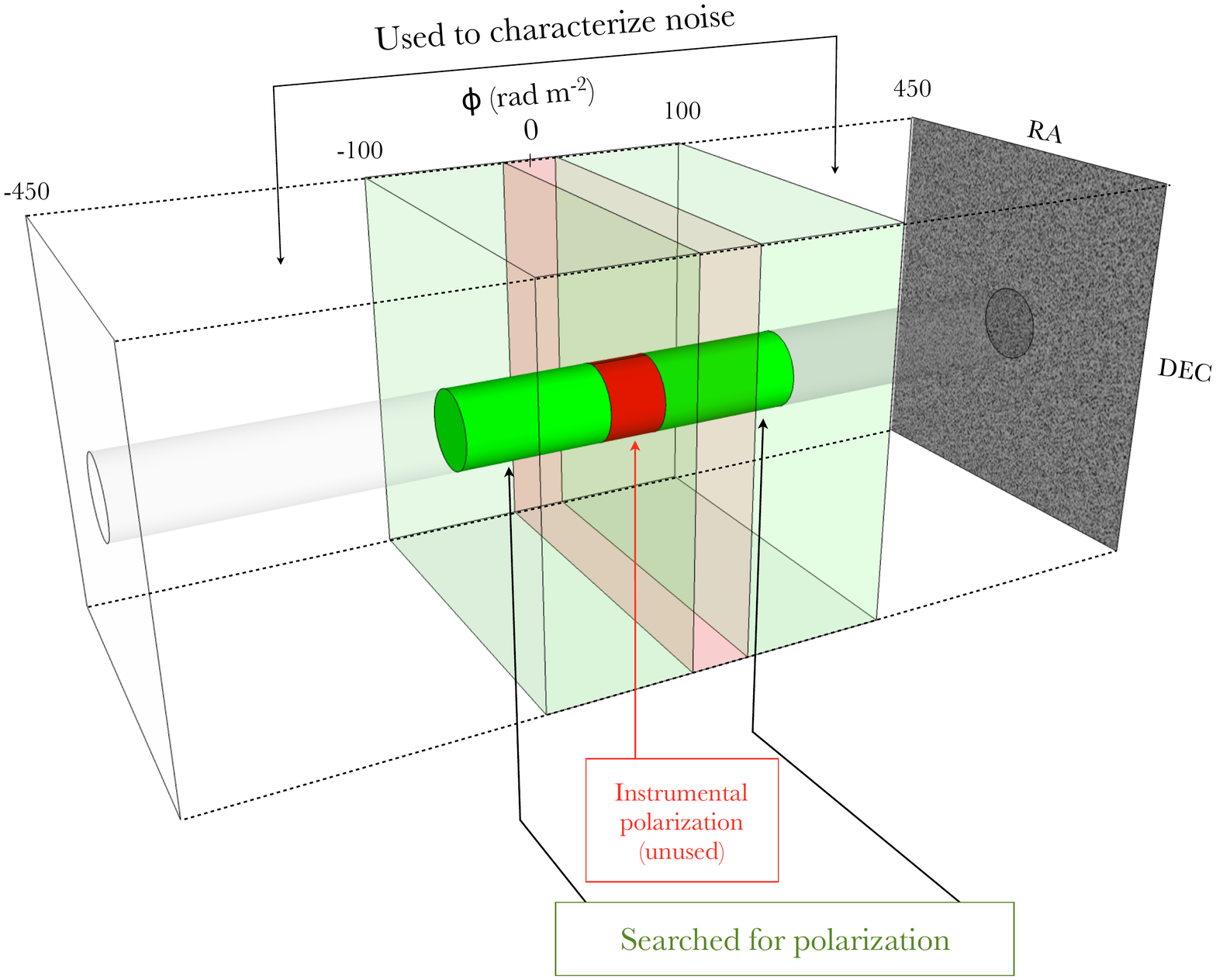}}
\end{figure*}

Figure~\ref{region_figure} is an illustration of the different regions used in the analysis. Those regions are listed below, and the criteria used to define their boundaries are explained.  

\begin{enumerate}
    \item The on-source and off-source regions. 
    \item The regions of high $|\phi|$, both on- and off-source, where no polarization is expected. 
    Those regions are used to characterize the noise. 
    \item The on-source region of low $|\phi|$ where polarization is searched for, excluding the region of instrumental polarization close to $\phi = 0$.
    \item The region of instrumental polarization close to $\phi = 0$. 
\end{enumerate}

To define a region that may contain polarization from the source, 
all pixels in (RA, Dec) with an intensity greater than a certain threshold, $I_{\rm thresh}$, were selected. 
\footnote{$I_{\rm thresh}$ was chosen such that $p_\mathrm{cell}( 0.05  I_\mathrm{thresh}) = 0.00135$, with $p_\mathrm{cell}$ defined in Sect.~\ref{source_pvals}. This means that the measurement of a region with a degree of polarization 5\% would have a $p$-value (introduced in Sect.~\ref{statistical_analysis}) of $0.00135$. 
With Gaussian noise, a signal at 3$\sigma$ would have this $p$-value.}     
Typically, $I_\mathrm{thresh}$ was of the order of $\unit[10]{mJy\,beam^{-1}}$.
Each source was inspected visually and the threshold was increased if artifacts (for instance due to phase errors) were seen. This was done for 28 sources.
    
In the Faraday depth dimension, we constrained our search to $|\phi|<\unit[100]{rad\,m^{-2}}$. 
The range around $\phi=\unit[0]{rad\,m^{-2}}$ required special attention 
because of the contamination by instrumental polarization. 
We always excluded $|\phi|\leq\unit[1.5]{rad\,m^{-2}}$ to exclude the instrumental peak itself. 
Additionally, instrumental polarization from the brightest sources creates artifacts at larger $|\phi|$ in the whole field. Therefore the standard deviation in each $\phi$-slice was measured (only including off-source pixels), creating a spectrum of the noise as a function of Faraday depth. The average and standard deviation of this Faraday spectrum at $|\phi|>\unit[20]{rad\,m^{-2}}$ were calculated, and we excluded the continuous range around $\phi=\unit[0]{rad\,m^{-2}}$ where the values were greater than five times the standard deviation above the average.

\subsection{Statistical analysis}
\label{statistical_analysis}

The key issue is to characterize the noise properties of the data in order to quantify the likelihood that a peak in polarized intensity observed in the Faraday cube is real. 
In the following subsections we define the different regions of interest, characterize the noise properties, and calculate the $p$-values of all the radio sources in our sample. 
The $p$-value (also sometimes called ``probability to exceed") is the probability of obtaining a value at least as high as the measured one in the absence of signal (that is, if the data contained only noise). 
The lower the $p$-value the higher the likelihood that the detection is real. 
In Sect.~\ref{fdr} we describe how the FDR method \citep{fdr_benjamini, fdr_miller} can be applied to quantify in a rigorous manner  
the fraction of false detections in a sample, based on the distribution of the $p$-values of the sources. 

\subsubsection{The null hypothesis: noise characterization}
\label{source_pvals}

We examined the distribution of local maxima at large Faraday depths  
($\unit[100]{rad\,m^{-2}}<|\phi|<\unit[450]{rad\,m^{-2}}$), 
    where it is assumed that no polarization is present. 
    
\begin{figure}
\centering
\includegraphics[width=\hsize]{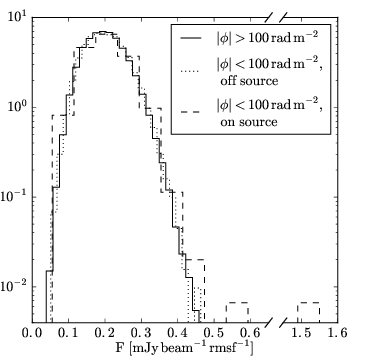}
\caption{Histograms of local maxima in different regions of the Faraday cube around the polarized source J132626+473741. The histograms have been normalized to facilitate comparisons. The instrumental polarization range, as defined in Section \ref{source_pvals}, has been excluded from the data. The distribution of local maxima on-source and at Faraday depths $|\phi|<\unit[100]{rad\,m^{-2}}$ shows a clear excess at larger $F$.
}
\label{noise_hist}
\end{figure}

 Figure~\ref{noise_hist} shows the distribution of local maxima in the Faraday cube of one of the sources in which polarization was found. On-source, the distribution of local maxima at $|\phi|<\unit[100]{rad\,m^{-2}}$ shows an excess of high polarization values.
 Off-source, no difference can be seen between the distributions at high and low Faraday depths. 
    
\begin{figure}[ht!]
\centering
{\includegraphics[width=8.8cm]{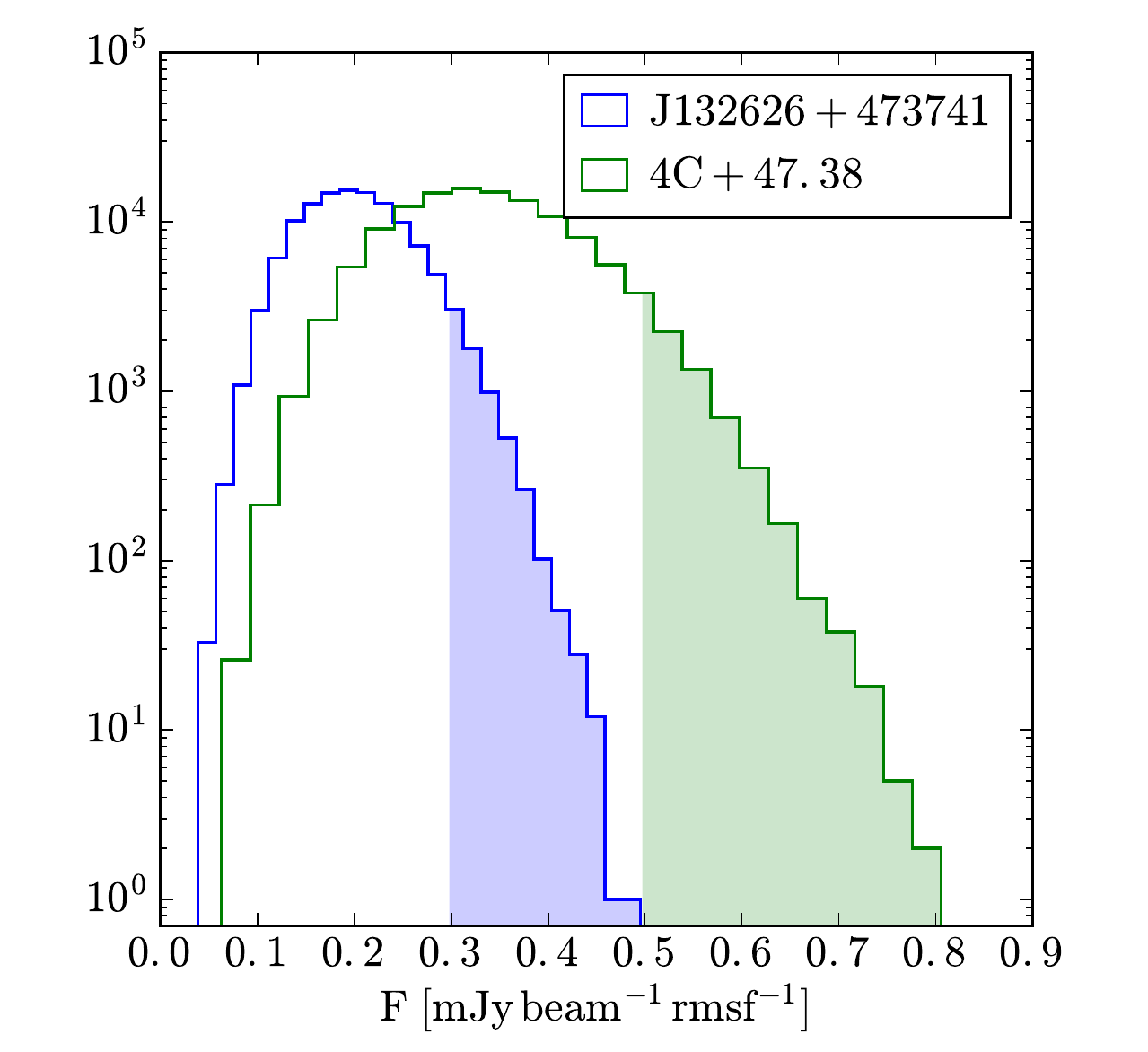}}
{\includegraphics[width=8.8cm]{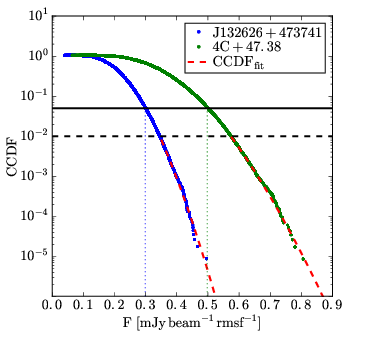}}
\caption{Characterization of the noise in the Faraday volumes of the two polarized sources, J132626+473741 (in blue, also presented in Fig.~\ref{noise_hist}) and 4C+47.38 (in green). 
The measurements were extracted in the range of Faraday depths $\unit[100]{rad\,m^{-2}} \leq |\phi |\leq \unit[450]{rad\,m^{-2}}$ where no polarized signal is expected. 
{\it Upper panel:} Histograms of local maxima. 
The histograms were not normalized since the two regions contain the same number of voxels. 
The shaded areas correspond to flux values in the top 5\% of the distribution. 
The noise in the Faraday cube of 4C+47.38 is higher than for the other source (the histogram is broader) because the source lies at a greater distance from the center of the field. 
{\it Lower panel:} Complementary cumulative distribution function (CCDF) for the same sources as in the first panel. 
 The dots show the actual CCDF and the solid line is the Gaussian fit used to model the distribution at high $F$
 (calculated at CCDF(F) $\leq 10^{-2}$, as indicated by the dashed line). 
 The horizontal solid line at CCDF = 0.05 corresponds to the lower limit of the shaded distributions in the upper panel.
 }
\label{R_of_P} 
\end{figure}
    
Figure~\ref{R_of_P} shows the complementary cumulative distribution (CCDF) of peaks that corresponds to the distribution of local maxima shown in Fig.~\ref{noise_hist}. 
Since we are interested in detecting polarized sources, which means identifying high values of $F$ that have a low probability of being due to noise, we need to quantify the distribution of the noise at high values of $F$ in regions where no polarized signal is expected. 
At high $F$, the CCDF of peaks can be well represented by a Gaussian. 
Therefore, we fit a Gaussian to the points at $CCDF (F) \leq 10^{-2}$ and use the fit as our CCDF at high values of $F$. 
The best-fit function is shown as a red dashed line in Fig.~\ref{R_of_P}. 

\subsubsection{Calculating the p-value of a source}

To calculate the $p$-value of a source, we searched for the highest peak (local maximum), $F_{\rm max}$, in the on-source region defined above. 
$p_\mathrm{cell}(F_\mathrm{max})$, is the probability of observing a peak at least as high as $F_\mathrm{max}$ in a given cell devoid of polarization. The $p$-value for the source, $p_\mathrm{source}$ is the probability of finding such a peak in \emph{any} cell. This probability is given by
\begin{equation}
    \label{psource_eq}
    p_\mathrm{source}=1- [1-p_\mathrm{cell}(F_\mathrm{max}) ] ^{N_{\rm cell}}\, ,
\end{equation}
where $N_{\rm cell}$ is the number of cells in the examined region. 

\subsubsection{The false discovery rate method}
\label{fdr}

Having calculated the $p$-value for each source, we used the FDR method \citep{fdr_benjamini, fdr_miller} to obtain a list of detected sources.

The FDR method allows one to select a number $\alpha$ in advance, and obtain a list of detections where the expected fraction of false detections is $\alpha$. The method works as follows:

The $p$-values are sorted in ascending order, and each is given an index $j$. Then the largest index is found for which 
\begin{equation}
\label{fdr_eq}
p_j < \frac{\alpha j}{N}
\end{equation}
\noindent where $N$ is the total number of measurements. All measurements with $p$-values smaller than $p_j$ are counted as detections. This can be understood intuitively by observing that $p_j N$ is the expected number of measurements with $p$-values below $p_j$, under the null hypothesis (i.e. false detections). $j$ is the actual number of measurements with such $p$-values. The proof is available in \citet{fdr_benjamini}.

\begin{figure}
\centering
\includegraphics[width=\hsize]{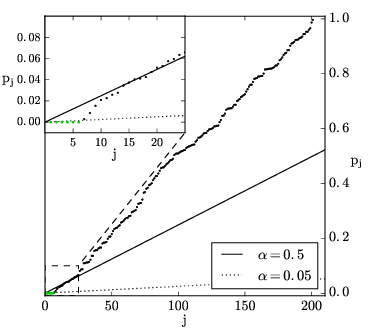}
\caption{Application of the FDR method to the 201 identified continuum sources around M~51 with two values of the false discovery rate, $\alpha$ (5\% and 50\%). 
Each dot shows the $p$-value of a source, $p_j$. The sources have been sorted by increasing $p$-value. 
The FDR method finds the (last) intersection of this distribution and a line with the slope $\frac{\alpha}{N}$, 
and classifies as reliable detections all the sources located to the left of the intersection (the green points have a 5\% false discovery rate). 
}
\label{fdr_explanation_fig}
\end{figure}

In our case, the total number of measurements is the number of examined radio continuum sources, $N = 201$. An illustration for two values of $\alpha$, 5\% and 50\%, is shown in Fig.~\ref{fdr_explanation_fig}. The green dots falling below the line that corresponds to $\alpha = 0.05$ correspond to the sources with a false discovery rate of 5\%. The results of the analysis are presented in the following Section. 

\section{Results}
\label{sectResults}

In Fig.~\ref{fdr_explanation_fig} we show the distribution of the examined sources sorted by increasing $p$-value. The inset shows more clearly the dots that fall below the lines that correspond to two values of the false discovery rate, $\alpha=0.05$ and $\alpha=0.5$. In 
Table~\ref{sources_table} we list the properties of those sources. 

\begin{table*} 
    \centering
    \caption{\label{sources_table} Most significant detections of polarization. 
    } 
    \begin{tabular}{lllllllll}
    \hline 
    \hline
         {\#} & {Name}   & {$\mathrm{R_{M\,51}}$} & {RA (J2000)} & { Dec (J2000)} & {I    } & {$\phi$}                 & {PI\tablefootmark{a}} & {$p$-value\tablefootmark{c}}\\
              &          &                        &              &                & {(mJy)} & {($\unit{rad\,m^{-2}}$)} & {(mJy)}               &  \\\hline
         \multirow{1}{*}{1.\tablefootmark{b}}& \multirow{1}{*}{J133920+464115} & \multirow{1}{*}{$1^\circ42'17''$} & \multirow{1}{*}{$13^\mathrm{h}39^\mathrm{m}23^\mathrm{s}$} & \multirow{1}{*}{$+46^\circ40'18''$} & \multirow{1}{*}{$3\,060 \pm 307$} & $+20.4 \pm 0.1$ & $3.0 \pm 0.11$ & \multirow{1}{*}{\ldots}\\
         
         2. & \multirow{1}{*}{4C+47.38} & \multirow{1}{*}{$2^\circ02'08''$}  & \multirow{1}{*}{$13^\mathrm{h}41^\mathrm{m}45^\mathrm{s}$} &       \multirow{1}{*}{$+46^\circ57'19''$} & \multirow{1}{*}{$5\,515 \pm 557$} & $+23.2 \pm 0.1$ & $3.6 \pm 0.13$ & \ldots\\
         
         3. & J132626+473741  & $42'50''$    & $13^\mathrm{h}26^\mathrm{m}32^\mathrm{s}$ & $+47^\circ37'58''$ & $507 \pm 51$     & $+3.0 \pm 0.1$ & $1.5 \pm 0.08$ & \ldots\\ 
         
         4. & J133707+485801   & $2^\circ08'43''$ & $13^\mathrm{h}37^\mathrm{m}08^\mathrm{s}$ & $+48^\circ58'03''$ & $1\,756 \pm 177$ & $+9.0 \pm 0.1$ & $1.8 \pm 0.15$ & \ldots\\ 
         
         5. & B3 1330+451  & $2^\circ21'28''$  & $13^\mathrm{h}32^\mathrm{m}47^\mathrm{s}$ & $+44^\circ53'35''$ & $705 \pm 71$    & $+14.0 \pm 0.1$ & $1.5 \pm 0.18$ & $4.3\times 10^{-7}$\\  
         
         \multirow{1}{*}{6.\tablefootmark{b}} & \multirow{1}{*}{J133613+490037} & \multirow{1}{*}{$2^\circ05'52''$} & \multirow{1}{*}{$13^\mathrm{h}36^\mathrm{m}16^\mathrm{s}$} & \multirow{1}{*}{$+49^\circ00'10''$} & \multirow{1}{*}{$561 \pm 56$}     & $+9.2 \pm 0.1$ & $1.1 \pm 0.15$ & \multirow{1}{*}{$1.3\times10^{-5}$}\\
    \hdashline
         7. & J133045+470318   & $12'18''$ & $13^\mathrm{h}30^\mathrm{m}45^\mathrm{s}$ & $+47^\circ03'19''$ & $119 \pm 12$ & $-98.0 \pm 0.1$ & $0.48 \pm 0.08$ & $0.0028$\\
         8. & J133051+475928   & $48'46''$ & $13^\mathrm{h}30^\mathrm{m}52^\mathrm{s}$ & $+47^\circ59'31''$ & $202 \pm 20$ & $+57.8 \pm 0.1$ & $0.49 \pm 0.08$ & $0.0086$\\
         9. & NGC 5256 (Mrk~266)   & $1^\circ46'51''$ & $13^\mathrm{h}38^\mathrm{m}18^\mathrm{s}$ & $+48^\circ16'41''$ & $585 \pm 59$ & $+1.8 \pm 0.1$ & $0.69 \pm 0.11$ & $0.015$\\
         10. & J133358+462204   & $1^\circ05'09''$ & $13^\mathrm{h}33^\mathrm{m}59^\mathrm{s}$ & $+46^\circ22'08''$ & $162 \pm 16$ & $-56.6 \pm 0.1$ & $0.53 \pm 0.09$ & $0.021$\\
         11. & B3 1323+476   & $44'04''$ & $13^\mathrm{h}25^\mathrm{m}47^\mathrm{s}$ & $+47^\circ26'09''$ & $881 \pm 89$ & $-57.0 \pm 0.1$ & $0.50 \pm 0.08$ & $0.023$\\
         12. & J132922+480239   & $51'09''$ & $13^\mathrm{h}29^\mathrm{m}22^\mathrm{s}$ & $+48^\circ02'41''$ & $478 \pm 48$ & $-5.0 \pm 0.1$ & $0.49 \pm 0.09$ & $0.026$\\
         13. & J132540+490955   & $2^\circ05'27''$ & $13^\mathrm{h}25^\mathrm{m}40^\mathrm{s}$ & $+49^\circ09'58''$ & $176 \pm 18$ & $+12.0 \pm 0.1$ & $0.84 \pm 0.15$ & $0.028$\\
         14. & B3 1330+459   & $1^\circ35'19''$ & $13^\mathrm{h}32^\mathrm{m}59^\mathrm{s}$ & $+45^\circ42'02''$ & $1311 \pm 133$ & $-4.4 \pm 0.1$ & $0.65 \pm 0.11$ & $0.034$\\
         15. & J133255+470046   & $33'03''$ & $13^\mathrm{h}32^\mathrm{m}56^\mathrm{s}$ & $+47^\circ00'49''$ & $211 \pm 21$ & $+16.6 \pm 0.1$ & $0.45 \pm 0.08$ & $0.037$\\
         16. & J132909+480107   & $49'54''$ & $13^\mathrm{h}29^\mathrm{m}09^\mathrm{s}$ & $+48^\circ01'09''$ & $347 \pm 35$ & $+10.6 \pm 0.1$ & $0.49 \pm 0.08$ & $0.040$\\
         17. & J133150+474557   & $39'36''$ & $13^\mathrm{h}31^\mathrm{m}51^\mathrm{s}$ & $+47^\circ46'00''$ & $137 \pm 14$ & $-41.2 \pm 0.1$ & $0.45 \pm 0.08$ & $0.041$\\
         18. & J133737+490439   & $2^\circ16'58''$ & $13^\mathrm{h}37^\mathrm{m}38^\mathrm{s}$ & $+49^\circ04'42''$ & $483 \pm 48$ & $+8.8 \pm 0.1$ & $0.94 \pm 0.16$ & $0.043$\\
         19. & B3 1324+473   & $29'24''$ & $13^\mathrm{h}27^\mathrm{m}03^\mathrm{s}$ & $+47^\circ05'46''$ & $423 \pm 42$ & $-56.0 \pm 0.1$ & $0.45 \pm 0.08$ & $0.047$\\
    \hline
    \end{tabular}
    \tablefoot{
    The 201 examined sources were those with a continuum flux density $S_{\rm 150~MHz} > 100$~mJy.
    In the top list of six sources, 5\% are expected to be false detections (i.e. less than one). 
    In the full table 50\% of the sources are expected to be false detections. 
    The sources are sorted by increasing $p$-value. 
    The names starting with a J are the names of the sources with counterparts in NVSS. The only exception is J132941.5+471734, that does not have any NVSS counterpart; the name comes from SDSS.
    Note that the coordinates listed here are those of the total-intensity source, not the exact location where a polarization peak was detected. \\
    \tablefoottext{a}{Due to the uncertainty of the polarization calibration, the calibration error has not been included.}\\
    \tablefoottext{b}{These sources have multiple significant Faraday peaks. Only the largest has been included in the table.}\\
    \tablefoottext{c}{A $p$-value given as \ldots means that it was too small for the numerical calculation.}
    }
\end{table*}

Setting $\alpha=0.05$ yields six polarized sources. 
The area covered is 19.6 deg$^2$. 
All of these sources were sufficiently polarized that the primary beam would not prevent detection anywhere within this region, and so the resulting detection rate is 1 source per 3.3 square degrees, or 0.3 source per square degree. 
These six sources are described individually in Sect.~\ref{sectDiscussion}. 

\begin{figure*} 
        \includegraphics[width=17cm]{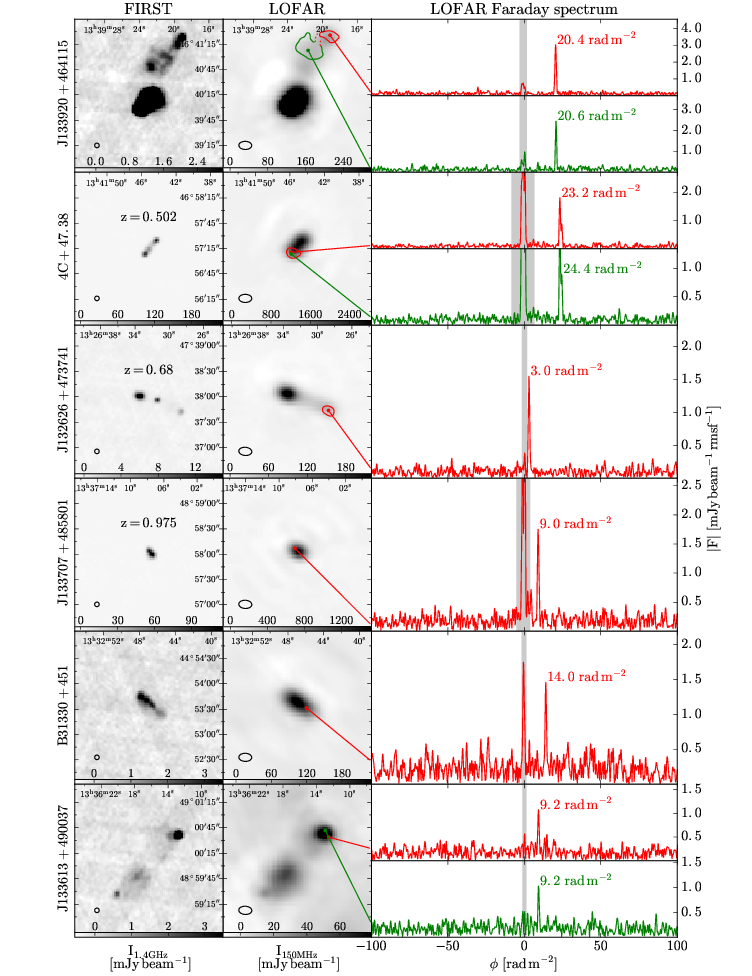} 
        \caption{The six sources detected in polarization in the LOFAR 150~MHz data with a 5\% false discovery rate. 
        {\it Left and middle column:} $3'\times3'$ VLA FIRST 1.4~GHz images (5$''$ resolution; \citealt{first95})  
        and LOFAR 150~MHz images. 
        The synthesized beams of the images are displayed in the bottom left corners. 
       {\it Right column:} LOFAR Faraday spectra at the most highly polarized location. 
        The contours correspond to the FWHM of the peak in polarized intensity; they were omitted when their shape was significantly affected by noise. 
        The red and green colors are used to show Faraday spectra at two nearby locations in the same source. 
        The grey shading around $\phi = 0$ shows the region of instrumental polarization that was excluded from the analysis (Sect.~\ref{subsectRegions}).
        } 
        \label{sources_fig}
\end{figure*}

Setting $\alpha=0.5$ gives 19 sources.  
Since half of the sources are expected to be false detections, this means than 9--10 sources are expected to be real. 
The list include the 6 most securely detected sources. 
The probability of having 
only six polarized sources in a sample of 
at least 19 sources with $\alpha= 0.5$ is only 8\%. 
This indicates that a few more sources (3--4) can be expected to be polarized. 
This brings the number density of polarized sources to about 0.5 per square degree. 

\begin{table*} 
   \caption{\label{depth_comparison} Sources with multiple Faraday depth measurements.}
   \centering
    \begin{tabular}{llllllll}
         \hline
         \hline
         & & {$\unit[150]{MHz}$} & {$\unit[150]{MHz}$} & {$\unit[610]{MHz}$} & {$\unit[1.4]{GHz}$} & {$\unit[1-2]{GHz}$} & {$\unit[1.4,1.6]{GHz}$}\\
 \# & Name   &  \multicolumn{1}{c}{This work}  & \citeauthor{mulcahy} & \citeauthor{farnes} & \citeauthor{taylor} & \multicolumn{1}{c}{\citeauthor{mao}} & \citeauthor{heald} \\ 
         &          & \multicolumn{1}{c}{\ }  & \citeyearpar{mulcahy} & \citeyearpar{farnes} & \citeyearpar{taylor} & \multicolumn{1}{c}{\citeyearpar{mao}} & \citeyearpar{heald}   \\ \hline 
    1.\tablefootmark{a} & J133920+464115& $ +20.4 \pm 0.1 $ & $ +20.5 \pm 0.1 $ & Outside FOV        &   $  +5.5 \pm  7.3 $ & Outside FOV    & Outside FOV \\ 
         2. & 4C+47.38    & $ +23.2 \pm 0.1 $ & $ +23.5 \pm 0.1 $ & Outside FOV        &   $ +30.6 \pm  1.4 $ & Outside FOV    & Outside FOV  \\ 
         3. & J132626+473741   & $ +3.0 \pm 0.1  $ & $  +3.2 \pm 0.1 $ & ND                 &   NI                & Outside FOV    & Outside FOV \\ 
         4. & J133707+485801    & $ +9.0 \pm 0.1  $ & $  +9.2 \pm 0.1 $ & Outside FOV        &   $   -8.9 \pm 3.2 $ & Outside FOV    & Outside FOV \\ 
6.\tablefootmark{a} & J133613+490037   & $ +9.2 \pm 0.1  $ &   ND              & Outside FOV        &   $ +11.1 \pm 10.6 $ & Outside FOV    & Outside FOV \\
         \hdashline
         7a. & J133045+470318  & $ -98.0 \pm 0.1 $&   ND                & $ -15.97 \pm 0.03$ &     ND               & $ +10 \pm 2 $ & Outside FOV \\ 
         8. & J133051+475928  & $ +57.8 \pm 0.1 $ &   ND                & Edge of FOV        &   $  -5.2 \pm 17.0 $ & Outside FOV   & Outside FOV \\
         11.& B3 1323+476   & $ -57.0 \pm 0.1 $ &   ND                & $ -16.97 \pm 0.03$ &     ND            & Outside FOV   & Outside FOV\\ 
         14.& B3 1330+459   & $  -4.4 \pm 0.1 $ & $-5.2 \pm 0.1 $     & Outside FOV        &     ND         & Outside FOV   & Outside FOV\\ 
\ldots\tablefootmark{b} & J132930+470612 & $ +96.2 \pm 0.1 $  &  ND                 & ND                 &     ND              & $ +21 \pm 2 $ & ND \\ 
         19.& B3 1324+473   & $ -56.0 \pm 0.1 $ &  ND                 & $ -8.11 \pm 0.07 $ &     ND            &  Outside FOV  & Outside FOV \\ 
         \hline
         \ldots& B3 1329+459 &  ND                 & $  -3.8 \pm 0.1 $ &     Outside FOV         & $ +35.0 \pm 16.3 $   & Outside FOV   & Outside FOV\\ 
         \ldots& B3 1326+470 &  ND                 &   ND                & $ -6.638 \pm0.013$ & $ +10.0 \pm 10.3 $   &   Outside FOV           & Outside FOV \\ 
         \ldots&  J132939+465909 & ND                  & ND                  & $ +11.15 \pm0.05 $ & $ -16.2 \pm 16.8 $   & $+16.6\pm0.3$ & $+14\pm1$\\ 
         \ldots& J133015+471026 &  ND                 &  ND                 & $ +33.52 \pm 0.03$ &    ND     & $ +26.0\pm0.4$& $+28\pm4$\\ 
        7b.  &  J133045+470318  &   ND                &  ND                 &$  -3.24 \pm 0.04 $ &  ND      & $ +17.2\pm0.8$ & $+17\pm2$\\ 
         52a  & J133124+471317 &   ND                &  ND                 & $ +11.51 \pm 0.03$ &   NI & $ +10.7\pm0.4$ & $+9\pm1$\\ 
         52b. & J133127+471300 &  ND                 & ND                  & $ +7.67 \pm 0.04 $ &   ND                & $ +6.0 \pm0.5$ & $+3\pm1$\\ 
         \ldots& B3 1331+472 &  ND                 &    ND               & $ +0.93 \pm 0.23 $ & $ +16.6 \pm 17.6 $   & Outside FOV       & Outside FOV\\ 
\ldots\tablefootmark{b} & J132941.5+471734 &  ND                 &  ND                 &  ND               &         ND             & $+23.5 \pm0.9$& $+20\pm1$\\ 
         \hline
    \end{tabular}
    \tablefoot{
       ND means that the source is not detected.
       NI means that it was not included in the \citet{taylor} RM catalog but detected in polarization in the NVSS catalog 
       of \citet{NVSS} with a polarized flux density (PI$_{\rm NVSS}$) greater than 3~mJy, which is below the 8$\sigma$ threshold to be included in the \citet{taylor} catalog.
    The sources are listed by increasing $p$-value.  
    The top list (above the horizontal dashed line) are the sources detected with a false discovery rate of 5\%. 
    Source~5 in Table~\ref{sources_table} is not included here because it was not listed in any of the other surveys. 
    The sources listed above the second line (including the top list) have a false discovery rate of 50\%. 
    The bottom list (below the horizontal line) contains the sources not detected by us but with a Faraday depth (or RM) measured in at least two other radio polarization studies. 
    7a and 7b are two components of the same sources but appear in different parts of the table. 
    52a and 52b are two components of the same source.\\
    \tablefoottext{a}{These sources have multiple significant Faraday peaks. Only the largest has been included in the table.}
    \tablefoottext{b}{These sources have a flux density at 150~MHz that is lower than 100~mJy.}
    }
\end{table*}

All the sources discussed above were part of the flux-density-limited sample ($S_{\rm 150 MHz} > 100$~mJy). We also imaged six sources fainter than $\unit[100]{mJy}$ that have been  detected in polarization in another radio frequency band. Of these, only one was detected (J132930+470612, with a $p$-value of $0.038$). This $p$-value is low enough for the source to be included in  $\alpha=0.5$ sample, but not in the top list with $\alpha=0.05$. To preserve the uniformity of the sample, this source is not included in Table~\ref{sources_table}, but it appears in Table~\ref{depth_comparison} where detections in different data sets are presented. 

The Faraday cubes of the 19 sources in the sample with a false discovery rate of 50\% will be published electronically.

\section{Discussion}
\label{sectDiscussion}

We start by comparing our measurements with those of \citet{mulcahy} that were based on the same data set. 
In Sect.~\ref{subsectSecureSources} we discuss the most securely detected sources (those with a FDR of 5\%) individually; 
we look at their morphology in the Faint Images of the Radio Sky at Twenty Centimeters (FIRST, \citealt{first95}) that have a higher angular resolution ($5''$) than the LOFAR images and search for optical counterparts and redshift estimates. 
Polarization measurements at other frequencies provide additional independent information that may help determine which ones of the sources in our second list (with an FDR of 50\%) are real. In Sect.~\ref{subSectOtherFreqs} we examine those measurements in more detail. 
In Sect.~\ref{sectPvalSNR} we compare the advantages of using $p$-values and the FDR method relative to using pre-defined signal-to-noise ratios. 
Many sources previously found to be polarized at higher frequencies \citep{farnes,mao,taylor,heald} are not detected at 150~MHz by LOFAR. 
This is to be expected, as depolarization is expected to be stronger at low frequencies \citep[e.g.][]{burn}. 
In Sect.~\ref{sectFaradayThick} we discuss the insensitivity of LOFAR to Faraday-thick sources. 

\subsection{Comparison with \texorpdfstring{\citet{mulcahy}}{Mulcahy et al. (2014)}}
\label{subsectCompaMulcahy}
Using the same LOFAR measurement set, \citet{mulcahy} had idenfied six polarized sources in the field 
using a pre-defined signal-to-noise threshold. A comparison of our two lists can be summarized as follows: 

\begin{itemize}

    \item There are four sources in common; they are the strongest detections 
and the measured Faraday depths are in very good agreement 
(see Tables~\ref{sources_table} and \ref{depth_comparison}). 

    \item Our fifth source (B3 1330+451) is outside the area searched by \citet{mulcahy}. 
    
    \item Our sixth source (J133613+490037) was not detected by \citet{mulcahy} but was detected at 1.4~GHz by \citet{taylor} (see Sect.~\ref{subsectSecureSources}). 

    \item The fifth source detected by \citet{mulcahy} (J133258+454201) appears in our longer list of 19 sources with $\alpha=0.5$ (B3 1330+459); the measured Faraday depths differ slightly between our two measurements ($-4.4\pm 0.1$~rad~m$^{-2}$ versus $-5.2\pm0.1$~rad~m$^{-2}$ for \citet{mulcahy}).
    
    \item \citet{mulcahy}'s sixth source (J133128+454002) is their most weakly polarized source. It did not make it into our list of sources with an estimated 50\% false discovery rate. 
    
\end{itemize}

\subsection{Sources detected with a 5\% false discovery rate}
\label{subsectSecureSources}

Let us examine more closely our most securely detected sources, i.e. the subsample with a false discovery rate of 5\%. In the left column of Fig.~\ref{sources_fig} we show images of the sources observed at higher angular resolution ($5''$) at 1.4~GHz by the VLA FIRST (Faint Images of the Radio Sky at Twenty Centimeters) Survey \citep{first95}. 
The LOFAR 150~MHz total-intensity images are shown in the middle column, and in the right column we show the Faraday spectra extracted from regions in which polarized emission was detected in the LOFAR data.

\textit{Source~1 (J133920+464115)} has a complex radio morphology. We find two regions of strong polarization in the northern part, peaking at $\unit[20.4\pm0.1]{rad\,m^{-2}}$ and $\unit[20.6\pm0.1]{rad\,m^{-2}}$, in agreement with what was found by 
\citet{mulcahy}. 
Like \citet{mulcahy}, who discussed the source as a radio galaxy as a core and a single lobe,  
we do not detect polarization from the bright ``core" at the center of the image. 
However, from the morphology of the high-resolution FIRST image it is not certain that the fainter features to the north are related to the ``core". There is no clear counterpart in SDSS, which suggests that the source(s) are distant or highly obscured.

\textit{Source~2 (4C+47.38; B3 1339+472)} is a double-lobed radio galaxy (only partially resolved with LOFAR) and the brightest source in the sample. Between the lobes there is a quasar with redshift $z=0.502\pm0.003$ \citep{b3_vla}. 
\citet{2003A&A...406..579K} 
derived an RM of $46.4\pm2.7$~rad~m$^{-2}$ from polarization measurements at 1.4, 2.7, 4.8 and 10.5~GHz, which is about twice as high as the Faraday depth that we measure in the LOFAR 150~MHz data. They also derived a spectral index of $-1.01$ between 408~MHz and 10.6~GHz. 

\textit{Source~3 (J132626+473741)} consists of three parts. The middle component is not visible in this observation, but can be seen in the FIRST image. A counterpart to the middle component was observed by the SDSS, with a redshift of $z=0.68240 \pm 0.000351$ \citep{sdss_z}.

\textit{Source~4 (J133707+485801)} is partly resolved as a double source in FIRST, but not by LOFAR. 
SDSS has an optical counterpart with a photometric redshift $z=0.975$ \citep{sdss_z_phot}.

\textit{Source~5 (B3 1330+451)} was not observed by \citet{mulcahy}, as it was outside their imaged field. It is only partially resolved by LOFAR. It is resolved into four sources by FIRST. 
The polarization detected by LOFAR is associated with the SW part. 

\textit{Source~6 (J133613+490037)} 
shows two distinct peaks in the northwest part, both at $\unit[9.2 \pm 0.1]{rad\,m^{-2}}$. 
Both parts of the source were detected in polarization at 1.4~GHz, 
with the southeast at an RM of $\unit[10.7 \pm 16.5]{rad\,m^{-2}}$ and the northwest at $\unit[11.1 \pm 10.6]{rad\,m^{-2}}$
\citep{taylor}. 
It was not detected in polarization by \citet{mulcahy}.

The six detected sources appear to be at least partially resolved by FIRST, and have a morphology consistent with that of double-lobed radio galaxies (Fig.~\ref{sources_fig}).

\subsection{Sources detected at other radio frequencies}
\label{subSectOtherFreqs}

Table~\ref{depth_comparison} lists the Faraday depths of the sources in the field that have been measured in at least two of the following studies:

\begin{itemize}
\item This work; 
\item \citet{mulcahy}: same calibrated LOFAR data as in this work, but analyzed differently; 
\item \citet{farnes}: 610 MHz GMRT observation of a fraction of the field; 
the polarization fraction was calculated for sources within a radial distance $\leq 35.6'$ from the center and should be considered as upper limits for the sources beyond a radial distance of $22.2'$ from the center; 
the full resolution of the data was $\sim5''$ and the analysis of the polarization was done on images at a resolution of $24''$; 
\item \citet{taylor}'s RM catalog is based on the NVSS survey \citep{NVSS} at 1.4~GHz with a resolution of $45''$. It covers the whole sky north of $-40^\circ$ and has an average  density of about one RM value per square degree. 
\item \citet{heald}: 1.4 and 1.6~GHz WSRT observations of the central part of the field ($34'\times34'$) 
at a resolution of $> 15''$. 
\item \citet{mao}: 1--2 GHz JVLA observations of the central part of the field ($40'\times40'$) 
at a resolution of $13.2''\times8.7''$. 
We noted typographical errors in one of the tables 
\footnote{The first column of Table~3 of \citet{mao} lists properties of polarized sources that are common to their study and to that of \citet{farnes}. 
The sources seem to be sorted in increasing values of RA, as in \citet{farnes}, but their names were extracted from \citet{mao}'s Table~2 where they had been listed in a different order. 
The correct order in the first column of Table~3 of \citet{mao} should be: J1329+4658c; J1330+4710; J1330+4703a; J1330+4703b; J1331+4713a; J1331+4713b.}. 
\end{itemize}

The fields of view of those observations are marked on Fig.~\ref{fig1}. 
Most of the sources that are listed in our Table~\ref{sources_table} are outside the fields of view of the targeted observations at higher frequencies. 
The RM catalog of \citet{taylor} covers the entire field, and we compare it with our detections in Sect.~\ref{compaTaylor}. 

\citet{mao} used a number of depolarization models to fit to their  polarization measurements in the 1--2~GHz band. In total, they modeled six sources (their Table~2;  since some of the sources had multiple components, a total of 10 components was modeled). Only one of their listed sources is detected in our study (our Source 7a; \citet{mao}'s source J1330+4703b). 
This source is particularly interesting because it was also detected at 610~MHz by \citet{farnes} and lies behind the prominent H{\sc i} tail of M~51. 
We discuss this source in Sect.~\ref{sectSource7}.

\subsubsection{Comparison with the \texorpdfstring{\citet{taylor}}{Taylor et al. (2009)} RM catalog}
\label{compaTaylor}

Of our six securely detected sources, four have an RM listed in \citet{taylor}'s catalog. 
We note that Source~3 was detected in polarization in the NVSS catalog ($3.08\pm0.69$~mJy), but below the 8$\sigma$ threshold to be included in the \citet{taylor} RM catalog. Source~5 was not clearly detected in the NVSS, with a polarized flux density of $0.83\pm0.51$~mJy. 
Of the two additional sources detected by \citet{mulcahy}, one has an RM entry in \citet{taylor}'s catalog; the other one, which coincides with our Source~14, has a polarized flux of only $0.81\pm0.40$~mJy in NVSS, so well below \citet{taylor}'s selection threshold. 
Of the 13 others that are included in our sample with a 50\% false discovery rate, only one source (our Source~8) figures in \citet{taylor}'s RM catalog.

This shows that detection of polarization at 1.4~GHz in the NVSS catalog or inclusion in \citet{taylor}'s RM catalog does not imply that the source may be detected in polarization in these 150~MHz LOFAR data. 
For the sources in common, there is no general agreement between the Faraday depths measured at 150~MHz and those measured by \citet{taylor} at 1.4~GHz. This might be due to resolution, sensitivity, and/or Faraday depolarization effects. 

\subsubsection{Galactic RM foreground}

M~51 is located at a high Galactic latitude ($b = +68.5^\circ$) where the rotation measure due to the Milky Way is expected to be low. From the five polarized sources in the field of their WSRT observations \citet{heald} estimated a foreground RM of $12\pm2$~rad~m$^{-2}$. 
\citet{mao} derived a median RM of 13~rad~m$^{-2}$ with a standard error of 1~rad~m$^{-2}$ from their JVLA 1--2 GHz measurements, 
excluding 
the sources located on sightlines with a neutral hydrogen column density larger than $10^{20}$~cm$^{-2}$ in the H{\sc i} map of \cite{1990AJ....100..387R}. 
For our entire field ($R_{\rm M51} < 2.5^\circ$), the mean RM of sources in the \citet{taylor} catalog is $12.0\pm 14.8$ rad~m$^{-2}$. 
The mean and standard deviation of RM values of the six polarized sources that are securely detected in the LOFAR data is $13.1\pm7.6$~rad~m$^{-2}$.
All those values are in agreement and provide an estimate of the Milky Way RM foreground in the direction of our observations. 

\subsubsection{Source~7: a radio source behind M~51's H{\sc i} tail}
\label{sectSource7}

\begin{figure*}
    \centering
    \includegraphics[trim={0 0.3cm 0 0},width=17.0cm]{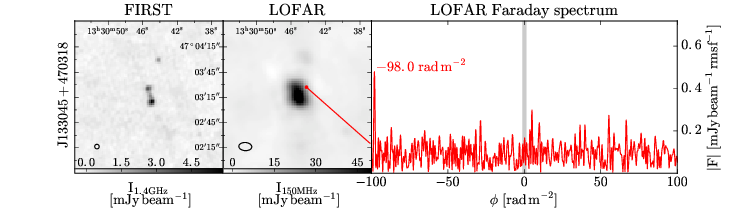}
    \includegraphics[width=8.8cm]{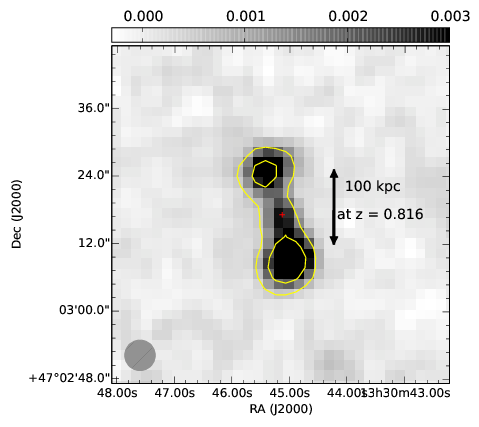}
    \includegraphics[width=8.6cm]{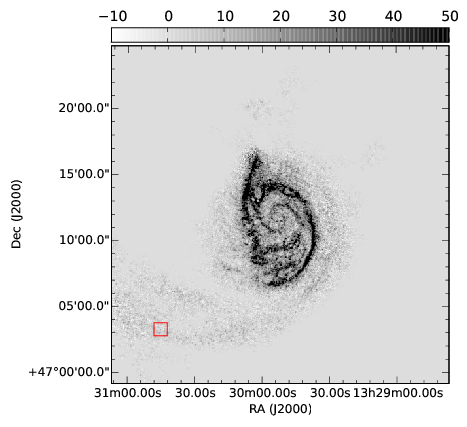}
    \includegraphics[trim={0 0.22cm 0 0.18cm},width=8.8cm]{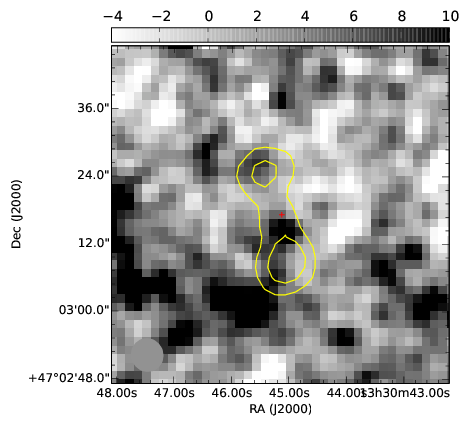}
    \includegraphics[width=8.8cm]{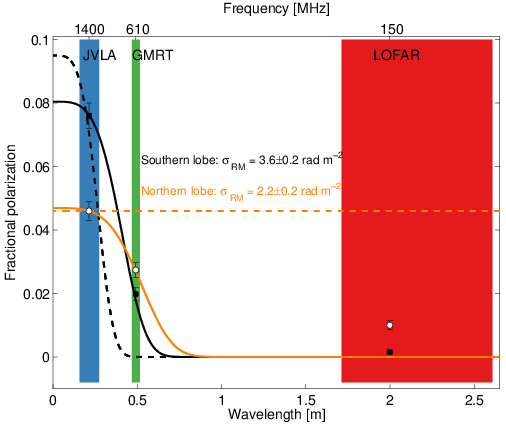}
    \caption{
    {\it Top row:} Same subfigures as in Fig.~\ref{sources_fig}, for Source~7. 
    The grey shading around $\phi = 0$ shows the region of instrumental polarization that was excluded from the analysis (Sect.~\ref{subsectRegions}). 
    {\it Second row, left panel:} $(1'\times1')$ FIRST 1.4~GHz image of our Source~7 in grey scale and in contours (1 and 3 mJy~beam$^{-1}$). 
    The red cross indicates the location of the optical counterpart to the radio core. 
    The color scale ranges from $-0.3$ to 3~mJy~beam$^{-1}$ and the angular resolution is 5$''$, as indicated by the grey circle in the bottom left corner.  
    {\it Second row, right panel:} H{\sc i} integrated intensity (moment 0) image of M~51 from THINGS. The red square has a size of $1'\times1'$ and indicates the location of Source~7, the radio galaxy displayed in the previous panel. 
    The grey scale is in Jy~beam$^{-1}$~m~s$^{-1}$.
    {\it Third row, left panel:} Zoom of the previous image: $(1'\times1')$ H{\sc i} high-resolution ($5.8''\times5.5''$) moment 0  image in grey scale.
    The yellow contours are the same radio contours of the FIRST image shown in the top left panel. 
    {\it Third row, right panel:} Measurements of the fractional polarization in the northern  
    radio lobe of Source~7 (white circles) and the southern lobe (black squares). 
    The dashed lines show the depolarization models that best fit the high-frequency data (JVLA 1--2 GHz, \citealt{mao}) and the solid lines depolarization models that match the averaged 1--2~GHz measurements and the GMRT 610~MHz of \citet{farnes}.  
    The LOFAR measurements are not used to constrain the models. 
    }
    \label{figSource7}
\end{figure*}

Source~7 lies 
at an angular distance of 12$'$ from the center of M~51 (or 26.5~kpc, assuming a distance to M~51 of 7.6~Mpc, \citealt{2002ApJ...577...31C}). 
This source source is of special interest because of its detection in polarization at several frequencies and its location behind the prominent tidal tail of neutral hydrogen 
discovered by \citet{1990AJ....100..387R} 
and imaged more recently by The H{\sc i} Nearby Galaxy Survey, THINGS \citep{2008AJ....136.2563W}. 
In the top right panel of Fig.~\ref{figSource7} we show the THINGS H{\sc i} integrated intensity image; 
the right square indicates the location of Source~7. 
No diffuse radio continuum emission of M~51 is detected in the area of the H{\sc i} tail \citep[e.g.][]{1992A&A...265..417H,2011MNRAS.412.2396F,mulcahy},  
as expected in the absence of cosmic-ray electrons outside the main star-forming disk of the galaxy. 
The tidal tail may, however, contain thermal electrons and magnetic fields that could cause Faraday rotation and/or depolarization from a background polarized source. 

In Fig.~\ref{figSource7} (top left panel) we show the image from the FIRST 1.4~GHz survey centered at the location of the radio source. The source has the morphological appearance of a double-lobed radio galaxy. 
Three sources are listed in the FIRST catalog:  
\begin{enumerate}
\item the rather faint core (with an integrated flux density of about 4~mJy; FIRST~J133045.1+470316), 
\item a northern component ($S_{1.4 {\rm GHz}} \simeq 9$~mJy; FIRST~J133045.3+470324), 
\item and a brighter southern component ($S_{1.4 {\rm GHz}} \simeq 11.8$~mJy; FIRST~J133045.0+470309) 
that is slightly more extended than the FIRST beam. 
\end{enumerate}
The combined flux of those three components is in excellent agreement with the NVSS flux measurement of $24.8\pm1.2$~mJy \citep{NVSS}, indicating that no extended emission is lost.
There is an optical counterpart to the radio core, SDSS~J133045.13+470317.2, marked by a red cross on Fig.~\ref{figSource7}, with a photometric redshift of $z = 0.816\pm0.0432$ (there is, however, a note in SDSS that the object's photometry may be unreliable). 
In the standard $\Lambda$CDM cosmology\footnote{$\Omega_m = 0.3$, $\Omega_\Lambda = 0.7$, $H_0  = 70$~ km~s$^{-1}$~Mpc$^{-1}$}, this gives a scale of 7.56~kpc~arcsec$^{-1}$. The distance between the northern and the southern radio components is $~15''$ ($\sim113$~kpc), with the northern component at a projected distance of about 60~kpc from the core and the southern one at about 56~kpc, on the plane of the sky.  
Those rather large distances suggest that both radio lobes are located outside the main halo of the host galaxy.  

\citet{mao} detected polarization in the 1--2~GHz band from both radio lobes, the southern one being more polarized ($7.6\pm 0.4$\%) than the northern one ($4.6\pm0.3$\%). 
Their best-fit model to the northern source is that of a simple uniform rotating Faraday screen, with a Faraday depth of $10\pm2$~rad~m$^{-2}$ and a constant fractional polarization of 4.6\%. 
This model {\it overestimates} the fractional polarization at 610~MHz, that was measured to be $2.74\pm0.24$\% 
(\citealt{farnes}\footnote{
The sources' two components are listed as \#9 and \#10 in \citet{farnes}'s Table~1, where Source~\#10 is the northern component, that corresponds to J1330+4703b of \citet{mao}.
}). 
The Faraday depth at 610~MHz is $-15.97\pm0.03$~rad~m$^{-2}$ \citep{farnes}, which is different from the value measured at 1--2~GHz. 

For the southern source, \citet{mao}'s best-fit model is a depolarizing Faraday screen (external Faraday dispersion, with  $\sigma_{\rm RM} = 8.4$~rad~m$^{-2}$). 
In this case, the model {\it underestimates} the polarization at 610~MHz, measured to be $1.98\pm0.02$\% \citep{farnes}. 
The Faraday depths are also different at the two frequencies: $17.2\pm0.8$~rad~m$^{-2}$ at 1--2~GHz and 
$-3.24\pm0.03$~rad~m$^{-2}$ at 610~MHz. 

What about the LOFAR measurements? The source has two components named 7a and 7b in our  Table~\ref{depth_comparison}. 
Our algorithm identifies a polarized signal in the overall region of the northern lobe in the LOFAR image 
(the radio source is barely resolved in LOFAR and the detection is slightly offset from the peak in total intensity, see first row of Fig.~\ref{figSource7}). 
Source~7a is not included in our top list of most securely detected sources but appears in the second list of sources detected in polarization with a false discovery rate of 50\%.
The measured Faraday depth is very large in absolute value (close to $-100$~rad~m$^{-2}$).   
The Faraday spectra is the area of the radio source contains a number of other peaks of similar strength at lower Faraday depths (in absolute value). 
For those reasons, we do not regard 
the measured level of polarized emission ($0.48\pm0.08$~mJy) as a robust measurement of polarization from the northern lobe. 
The algorithm does not find any polarization towards the southern lobe. 

To calculate the degree of polarization of each lobe at 150~MHz, 
we assume that the core has a constant a flux density of 5~mJy 
(in agreement with the observations at higher frequencies), and estimate the flux density of each lobe by assuming that the lobes have the same flux density ratio as at 1.4~GHz. 
The total flux density is $S_{150~{\rm MHz}}^{\rm total source} = 119\pm12$ mJy. 
This gives about 49~mJy for the northern lobe and 65~mJy for the southern one,  
and a fractional polarization of 1\% for the northern lobe, and an upper limit of 0.15\% for the southern lobe, adopting a limit on the polarized emission of 0.1~mJy. 

From the measurements at 1--2~GHz by \citet{mao} and at 610~MHz by \citet{farnes}, we can use a simple depolarization model of an external Faraday screen to calculate the Faraday dispersion. 
For the northern lobe, we obtain 
$\sigma_{\rm RM} = 2.2\pm0.2$~rad~m$^{-2}$,
and for the southern lobe 
$\sigma_{\rm RM} = 3.6\pm0.2$~rad~m$^{-2}$. 
In the last panel of Fig.~\ref{figSource7} we show the measurements, the depolarization models of \citet{mao} (dashed lines), and the depolarization models by external Faraday dispersion derived from the averaged 1-2~GHz and the 610~MHz measurements (solid lines). 
The 1--2~GHz measurements gave a significantly larger Faraday dispersion for the southern lobe and therefore a stronger depolarization at longer wavelengths. 
Given the uncertainty of the LOFAR detection towards the northern lobe, we refrain from using this measurement to constrain the nature of the depolarization. However, if the LOFAR detection towards the northern lobe is real, then the fractional  polarization would decrease less steeply with wavelength than in the \citet{burn} 
model. It woud be more compatible with the model for external Faraday dispersion discussed by \citet{1991MNRAS.250..726T} 
that decreases as the inverse of $\sigma_{\rm RM} \lambda^2$ at long wavelengths, as found in other low-frequency observations of polarized sources \citep[e.g.][]{2013A&A...559A..27G}. 

The high-resolution ($\sim 5''$) H{\sc i} image of the region shows some substructure, with an H{\sc i} peak in M~51 south of the radio core of the background source (bottom left panel of Fig.~\ref{figSource7}). There is little neutral hydrogen, however, at the location of the radio lobes. We also examined the corresponding first and second moment images (velocity field and velocity dispersion) from THINGS and did not find any clear evidence of regular or turbulent velocity flows in that region that may have helped interpret the measurement (the very large Faraday depth in absolute value seen by LOFAR and the stronger Faraday dispersion towards the southern lobe). 

In the future, more sensitive broad-band polarization measurements and H{\sc i} observations with higher surface brightness sensitivity may make it possible to investigate in more details the magneto-ionic medium in the outer regions of galaxies via Faraday tomography of background sources. 
The VLASS\footnote{\tt https://public.nrao.edu/vlass/}, 
MeerKAT's MIGHTEEpol 
and MHONGOOSE 
surveys \citep{2017arXiv170901901J, 
2017arXiv170908458D} 
and ASKAP's WALLABY\footnote{\tt http://www.atnf.csiro.au/research/WALLABY/} 
and POSSUM\footnote{\tt http://askap.org/possum/Main/HomePage} surveys will give just these improvements. 

\subsection{p-values versus signal-to-noise ratios}
\label{sectPvalSNR}
A natural question is whether the method presented here has clear advantages over the traditional "sigma clipping" method based on selecting sources above of a pre-selected signal-to-noise ratio (S/N), where the noise is usually taken as the standard deviation of a distribution of measurements that are expected to be free from signal. 

The two approaches have several steps in common. In particular, the regions expected to be signal-free have to be defined. In the case of Gaussian statistics, the two methods are strictly equivalent. For a Gaussian distribution of noise, the significance of a given peak in the data cube can be uniquely quantified by its S/N or its $p$-value. 
A non-Gaussian noise distribution cannot, in general, be uniquely described by its standard deviation. In that case, looking at the $p$-values might be more relevant as it gives the probability of having a peak of a certain strength given the underlying noise distribution. The stronger the source the lower the probability of having a similar one in the noise distribution, and therefore the actual noise distribution has to be modeled, based on the weaker data points, and extrapolated to high values, as we did in Sect.~\ref{source_pvals}. 

The noise distribution of the polarized intensity in the Faraday cubes is clearly non Gaussian. Nevertheless, we calculated the standard deviation and the S/N for the investigated radio continuum sources. We found that the six securely detected sources all have a S/N greater than 10. For the others, there is a clear anticorrelation between the S/N and $p$-values, as expected, but the relation has a large scatter. 

When choosing selection criteria for larger surveys one must always balance the power of the survey against the possibility of false detections. Erring on the side of inclusion and assigning a $p$-value to each source in the resulting catalog gives users the ability to modify this balance to suit their own needs. The fraction of false detections is an intuitive parameter to use for this purpose.

\subsection{Insensitivity to Faraday-thick sources?}
\label{sectFaradayThick}
The LOFAR data set used in this study has a very small maximum scale in Faraday depth ($\Delta\phi_{\rm max}\sim 1$~rad~m$^{-2}$). This means that the measurements are partly insensitive to Faraday-thick sources. The Faraday spectra, in effect, pass through a high-pass filter. At these low frequencies, this effect is so large that we do not observe the total polarized emission, but instead steep gradients in emission with respect to Faraday depth. 
Differences in  the maximum Faraday range that can be detected by different instruments might explain the non-detection of some of the sources that were detected at other wavelengths. For instance, the 1--2 GHz VLA observations of \citet{mao} have a poorer resolution in Faraday space (the FWHM of their RMSF is 90~rad~m$^{-2}$), but they are more sensitive to extended structure in Faraday space (with a 50\% sensitivity to Faraday extents of 118~rad~m$^{-2}$). The different Faraday depths observed at different wavelengths may be an indication that many sources are Faraday complex.

Faraday-thick emission with internal structure on scales of $\unit[1]{rad\,m^{-2}}$ or less would show up as several smaller peaks. While any given peak might be too small to be detected confidently, the on-source Faraday spectra would still be busier than the surroundings. 
``Busy spectra" can be due to turbulent magnetic fields, as shown in the model by \citet{2012A&A...543A.113B} 
(see their Fig.~3, and what LOFAR can detect, their Figs.~8 and 9).
Our method might be extended to deal with multiple Faraday peaks. One could, for example, calculate a $p$-value using the few highest peaks. A positive correlation between the noise of adjacent cells may, however, cause false positives.

\section{Summary and conclusion}
\label{sectConclusion}

We have developed a new method to identify polarized sources in radio continuum data. 
We calculate the $p$-values of sources in Faraday cubes and use the FDR method to construct a list of polarized sources, of which a preselected fraction are expected to be false detections. 
We applied this method to the LOFAR observations of the M~51 field and confidently identified six sources, giving a number density of 1 polarized source per 3.3 square degrees at 150~MHz, or 0.3 source per square degree. 
The number density increases to 0.5 per square degree taking into account the larger sample of 19 sources with a FDR of 50\%. 

All six most secure detections are associated with radio sources that have multiple radio components and/or diffuse continuumm emission. Their morphology is consistent with that of double-lobed radio galaxies and in some cases the polarization comes from the outer lobes.
Correlation of our Table~3 (19 sources) with the catalog of double-lobed radio sources identified in the FIRST survey by \citet{2015MNRAS.446.2985V} 
gave 10 matches (50\%); cross-correlation of the whole catalog of 201 sources gave 83 matches (41\%). 
This indicates that a significant fraction of the sources that are polarized at low frequencies are classical radio galaxies, possibly of FR~{\sc ii} type \citep{1974MNRAS.167P..31F}. 
This is also found in the polarization study by \citet{2018A&A...613A..58V} 
of part of the LOFAR Two-meter Sky Survey \citep{2017A&A...598A.104S} 
that resulted in a catalog of 92 polarized sources. 
Low-frequency observations of polarization and Faraday rotation bring valuable information to constrain the properties of radio galaxies and their surroundings 
\citep[e.g.][]{2018MNRAS.475.4263O}). 

Our search was done on data from one LOFAR field image at a resolution of $18''\times15''$, while 
\citet{2018A&A...613A..58V} 
surveyed a much larger area at lower angular resolution ($\sim 4'$). The M~51 field is included in the data set 
used by \citet{2018A&A...613A..58V}, 
and only one polarized source was found (the brightest one in our sample).  

Our pipeline is well suited to LOFAR data but could be applied to other radio polarization measurements (e.g. MeerKAT, POSSUM, SKA-low and SKA-mid). 
Imaging individual sources at higher resolution and analyzing the corresponding RM cubes is computationally more efficient than dealing with very large cubes; 
it makes it possible to identify a larger number of polarized sources and quantify the rate of false detections. 

In the future, we intend to apply the method to more polarization data sets, in particular fields from 
the LOFAR Two-meter Sky Survey of the entire northern hemisphere, LOTSS, 
and the deep ongoing LOFAR observations of the GOODS-North field\footnote{PI Anna Scaife}.

Future improvements will include the identification of several peaks in the Faraday spectra (not only the strongest one), since a number of sources seem to have a complex Faraday spectrum. Depolarization may affect different Faraday components differently and might be the reason why the Faraday depths that are measured at low radio frequencies sometimes differ from the ones measured at higher frequencies. 

\begin{acknowledgements}
LOFAR, the Low Frequency Array designed and constructed by ASTRON, has facilities in several countries, that are owned by various parties (each with their own funding sources), and that are collectively operated by the International LOFAR Telescope (ILT) foundation under a joint scientific policy. 
This work made use of THINGS, `the H{\sc i} Nearby Galaxies Survey' \citep{2008AJ....136.2563W}. 
This research has made use of the NASA/IPAC Extragalactic Database (NED) which is operated by the Jet Propulsion Laboratory, California Institute of Technology, under contract with the National Aeronautics and Space Administration.
This research has made use of the VizieR catalogue access tool, CDS, Strasbourg, France. The original description of the VizieR service was published in A\&AS 143, 23.
We have also made use of the table analysis software {\tt topcat} \citep{topcat}.
This research made use of {\tt Astropy}, a community-developed core Python package for Astronomy \citep{astropy}.
This research also made use of the Matplotlib plotting library \citep{matplotlib}. 
We thank Aritra Basu for useful comments on the manuscript.  
We thank Steven Black Design for producing Fig.~\ref{region_figure}.
D.D.M. gratefully acknowledges support from ERCStG 307215 (LODESTONE).
C.H. and A.F. gratefully acknowledge support from the Gothenburg Centre for Advanced Studies, GoCAS. AF also thanks STFC and the Leverhulme Trust for support. 
We thank the referee for pointing out interesting references and for other useful comments.
\end{acknowledgements}

\bibliographystyle{aa}
\bibliography{main.bib}
\end{document}